\documentclass[12pt,a4paper,final]{iopart}

\usepackage{iopams}  
\usepackage{graphicx}

\usepackage{dcolumn}
\usepackage{bm}
\usepackage{hyperref}

\RequirePackage{cleveref}

\newcommand{\ket}[1]{\ensuremath{|#1\rangle}}
\newcommand{\bra}[1]{\ensuremath{\langle #1|}}
\newcommand{\unit}[2]{\ensuremath{#1}\hspace{2pt}{#2}}
\newcommand{\Mg}[1]{\ensuremath{^{#1}}Mg\ensuremath{^{+}}}
\def\dsoh{\fm{{}^2\mathrm{S}_{1/2}}\xspace}
\def\dpoh{\fm{{}^2\mathrm{P}_{1/2}}\xspace}

\begin{document}

\title[Detection of motional ground state population of a trapped ion using delayed pulses]{Detection of motional ground state population of a trapped ion using delayed pulses}

\author{F. Gebert$^1$, Y. Wan$^{1,2}$, F. Wolf$^1$, Jan C. Heip$^1$, and Piet O. Schmidt$^{1,3}$}
\address{$^1$Physikalisch-Technische Bundesanstalt, 38116 Braunschweig, Germany}
\address{$^2${\it Present address:} National Institute of Standards and Technology, Boulder, Colorado 80305-3328, USA}
\address{$^3$Institut f\"ur Quantenoptik, Leibniz Universit\"at Hannover, 30167 Hannover, Germany}

\ead{Piet.Schmidt@quantummetrology.de}

\begin{abstract}

Efficient preparation and detection of the motional state of trapped ions is important in many experiments ranging from quantum computation to precision spectroscopy. We investigate the stimulated Raman adiabatic passage (STIRAP) technique for the manipulation of motional states in a trapped ion system. 
The presented technique uses a Raman coupling between two hyperfine ground states in \Mg{25}, implemented with delayed pulses, which removes a single phonon independent of the initial motional state. We show that for a thermal state the STIRAP population transfer is more efficient than a stimulated Raman Rabi pulse on a motional sideband. In contrast to previous implementations, a large detuning of more than 200 times the natural linewidth of the transition is used. This approach renders STIRAP suitable for atoms in which resonant laser fields would populate fluorescing excited states and thus impede the STIRAP process.
We use the technique to measure the wavefunction overlap of excited motional states with the motional ground state. This is an important application for photon recoil spectroscopy and other force sensing applications that utilize the high sensitivity of the motional state of trapped ions to external fields. Furthermore, a determination of the ground state population enables a  simple measurement of the ion's temperature.
\end{abstract}

\pacs{03.75.Be, 32.80.Qk, 42.50.-p}
\vspace{2pc}
\noindent{\it Keywords}: Adiabatic state manipulation, STIRAP, trapped ions, motional state population
\submitto{\NJP}
\maketitle

\section{\label{sec:level1} Introduction}
Progress in trapped-ion quantum information processing \cite{leibfried_experimental_2003, haffner_quantum_2008, blatt_entangled_2008, wineland_nobel_2013}, quantum simulation \cite{schaetz_focus_2013, blatt_quantum_2012}, and precision spectroscopy experiments \cite{schmidt_spectroscopy_2005, rosenband_frequency_2008, hempel_entanglement-enhanced_2013, huntemann_improved_2014,wan_precision_2014,gebert_precision_2015} is largely based on advances in the ability to control and manipulate the quantum states of the system. Trapped and laser-cooled ions represent a particularly well-controlled system for which different techniques have been established to control the internal (electronic) and external (motional) state. Commonly, sequences of laser or microwave pulses are applied to prepare a desired state or implement operations for state manipulation. For this, mostly square pulses with a fixed length and frequency are employed 
that rotate the atomic qubit and -- depending on the experimental implementation -- also change the motional state. The effect of undesired frequency components in square-shaped pulses has previously been reduced by employing amplitude-shaped pulses with a smooth rising and falling slope \cite{riebe_process_2006, benhelm_towards_2008, kirchmair_deterministic_2009}.
Furthermore, composite pulses, first developed in the context of nuclear magnetic resonance \cite{levitt_nmr_1979, levitt_composite_1986, vandersypen_nmr_2005}, are used in trapped ion systems to implement complex algorithms \cite{gulde_implementation_2003, schmidt-kaler_realization_2003} or operations that are less sensitive to variations of the experimental parameters \cite{timoney_error-resistant_2008, ivanov_high-fidelity_2011, shappert_spatially_2013, mount_error_2015}. 

Adiabatic state manipulation represents another class of techniques with reduced sensitivity to fluctuations in the coupling strength \cite{bergmann_coherent_1998, bergmann_perspective:_2015}. Pulses with slowly varying intensity and/or frequency are used to manipulate the state of the system. For trapped ions two adiabatic techniques have been investigated, namely Rapid Adiabatic Passage (RAP) and stimulated Raman adiabatic passage (STIRAP). In RAP a frequency and amplitude modulated pulse is used to tailor the dynamics of the atomic state dressed by the light field for adiabatic transfer of population between two bare atomic states. In the experiment, the time dependence of the intensity usually has a Gaussian shape, whereas the frequency is varied linearly in time across an atomic resonance. RAP has been used in optical qubits on carrier transitions for robust internal state preparation \cite{wunderlich_robust_2007, yamazaki_robust_2008, noel_adiabatic_2012, poschinger_interaction_2012}, and on sideband transitions, to prepare Fock \cite{watanabe_sideband_2011} and Dicke \cite{linington_robust_2008, toyoda_generation_2011} states. The STIRAP technique is typically realized in $\Lambda$-systems and relies on an adiabatic evolution from an initial to a final state without populating a short-lived intermediate state. It is usually implemented using Gaussian-shaped intensity profiles of two laser pulses with a fixed frequency difference that are delayed with respect to each other in time. It has been demonstrated for population transfer \cite{sorensen_efficient_2006} and the generation of Dicke states \cite{noguchi_generation_2012}, and suggested for efficient qubit detection of single ions \cite{moller_efficient_2007} and Doppler-free efficient state transfer in multi-ion crystals \cite{kamsap_coherent_2013}. 

Here we demonstrate STIRAP between hyperfine qubit states in \Mg{25} involving a change in the motional state. The coupling strength of such sideband transitions is strongly dependent on the initial motional state of the ion \cite{wineland_experimental_1998}.
We used the insensitivity of STIRAP to the coupling strength to perform a complete population transfer of motionally excited states to determine the motional ground state population.
For thermal states the ground state population is a direct measure for the temperature \cite{wan_efficient_2015}. Using this approach, good agreement with the expected Doppler cooling temperature is found.

We implement STIRAP using a large detuning, which is in contrast to the near resonant STIRAP transfer typically discussed in the literature \cite{bergmann_coherent_1998, fewell_coherent_1997}. In this situation the counter-intuitive and intuitive pulse sequences give comparable population transfer efficiency, allowing the pulse order to be chosen in order to minimize off-resonant scattering. The intuitive pulse sequence was previously studied in doped crystals and termed b-STIRAP \cite{klein_robust_2007}. The comparable large detuning used in our experiment also relaxes the condition of adiabaticity during the transfer process and consequently allows the transfer to be comparably fast. Furthermore, spontaneous emission from light fields coupling to states not involved in the STIRAP process is suppressed, allowing STIRAP to be implemented in multi-level systems such as the \Mg{25} ion used in our work. 

The paper is organized as follows. In \cref{sec:level2} we provide an introduction into the theoretical treatment of STIRAP. The experimental setup for the realization of STIRAP with a single trapped \Mg{25} ion is briefly described in \cref{sec:level3}. The implementation of numerical simulations supporting the experimental findings is described in \cref{sec:level4}. In \cref{sec:level5} we present the experimental results of our investigation on the STIRAP efficiency and its dependence on pulse order, pulse length and pulse separation together with the numerical simulations. An optimized pulse sequence is used to demonstrate the advantage of STIRAP over a stimulated Raman Rabi population transfer on carrier and sideband transitions for a thermal state. \Cref{sec:level6} summarizes the work and points at possible improvements and applications of the technique.

\section{\label{sec:level2} Principles}

\begin{figure}
	\centering
		\includegraphics[width=0.80\textwidth]{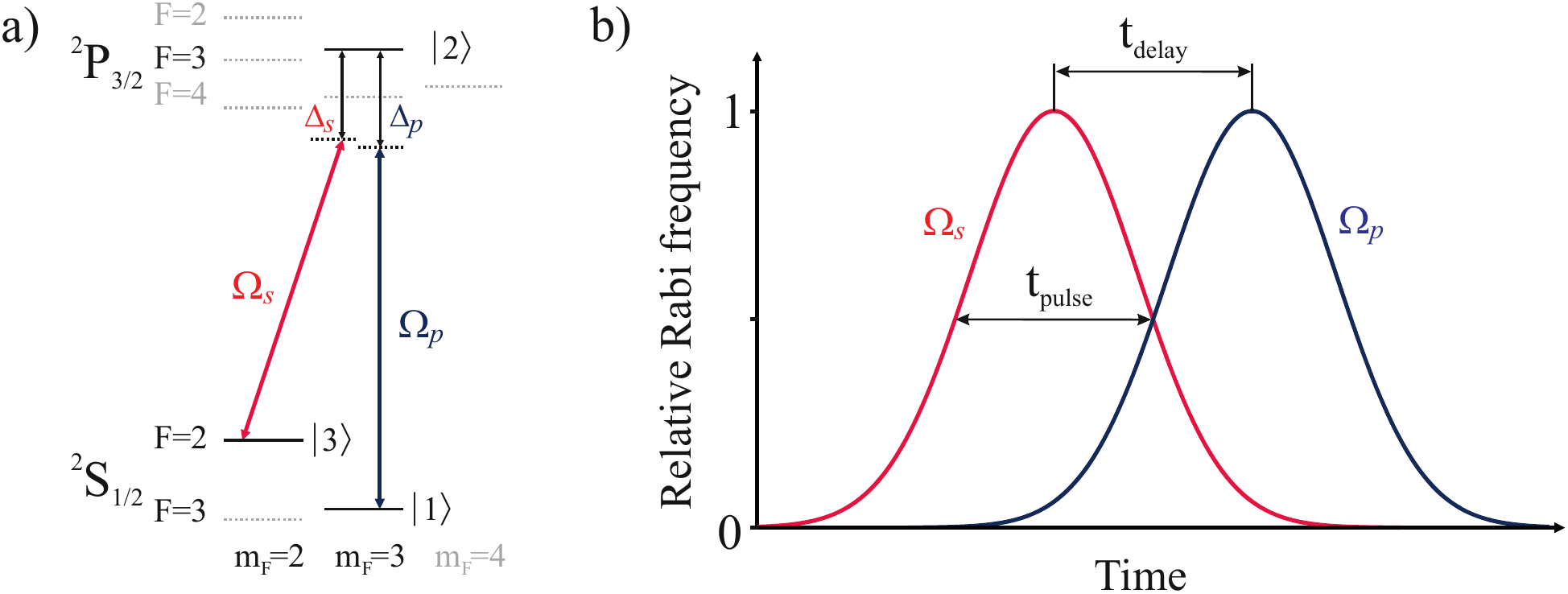}
	\caption{a) Simplified level structure of \Mg{25} together with the involved laser couplings with associated Rabi frequencies $\Omega_p$ and $\Omega_s$. The dark levels are relevant for the STIRAP process, whereas the grey, dotted levels lead to additional off-resonant couplings. b) Time dependence of the Rabi frequencies normalized to their maximum value. The pulse length is defined as the full width at half maximum and the delay of the two pulses is defined as the separation of the Rabi frequency maxima.}
	\label{fig:3lvl_pulses}
\end{figure}
\begin{figure}[tbp]
\centering
\includegraphics[width=0.32\textwidth]{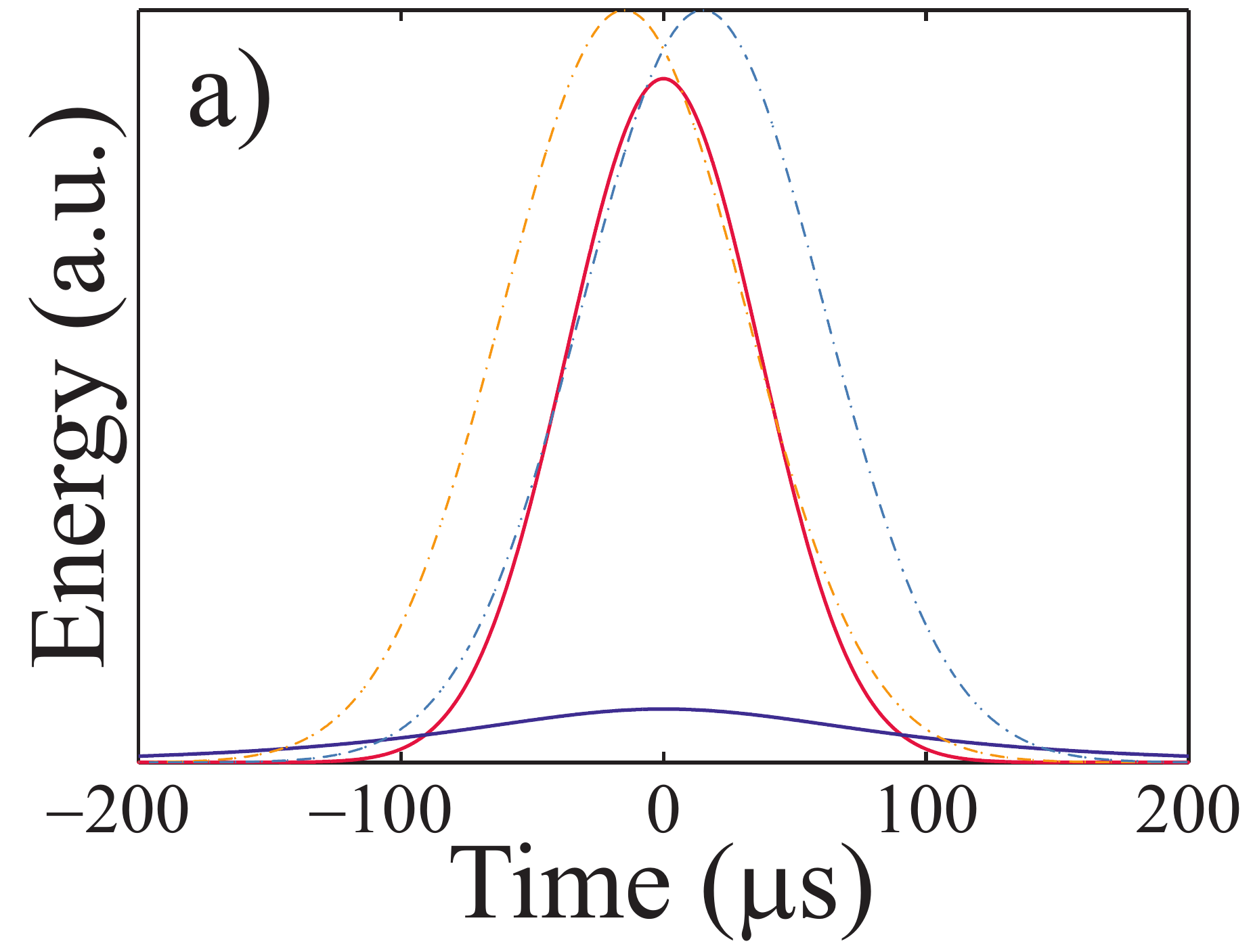}
	\label{fig:ac_timedep_sf03}
\includegraphics[width=0.32\textwidth]{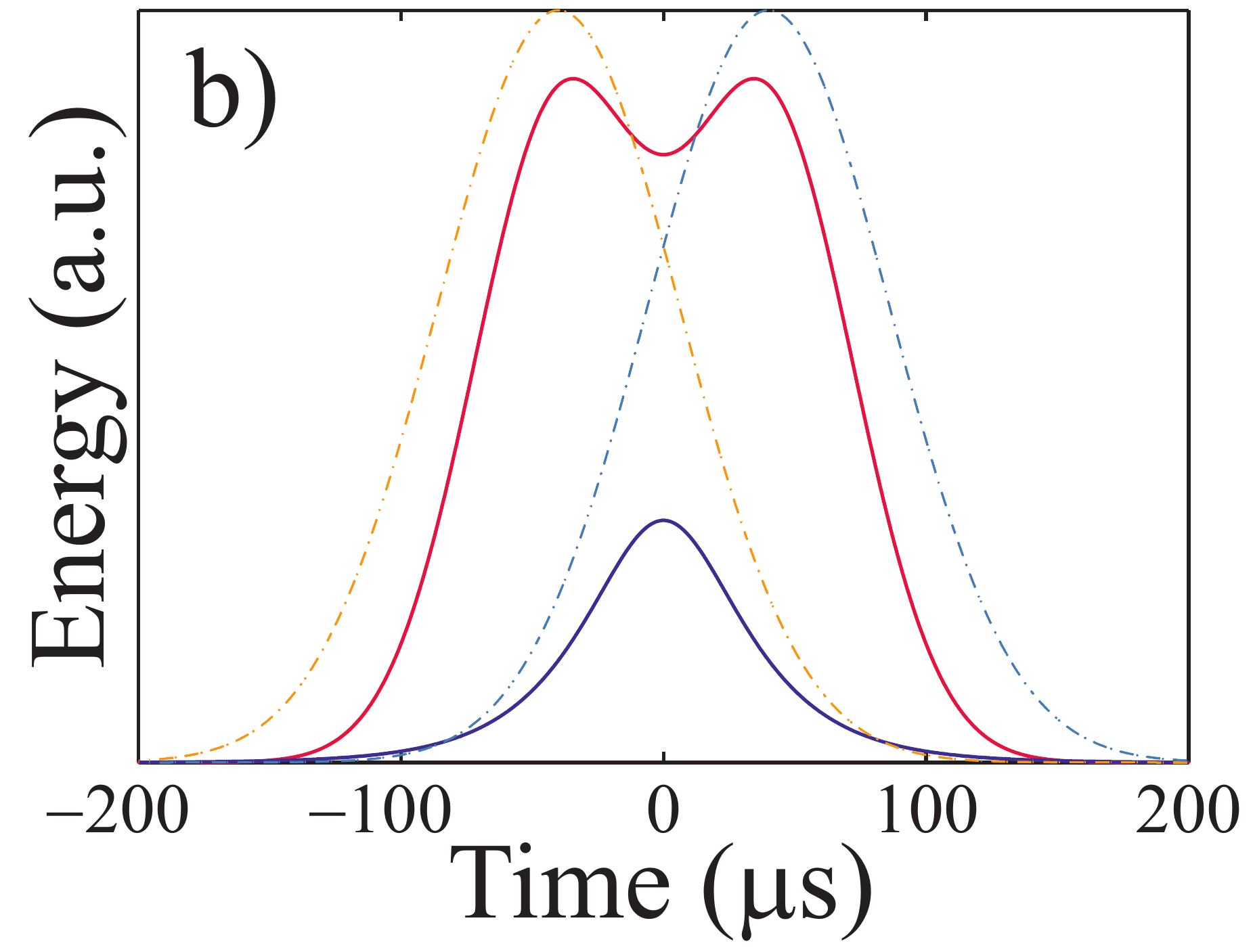}
	\label{fig:ac_timedep_sf08}
\includegraphics[width=0.32\textwidth]{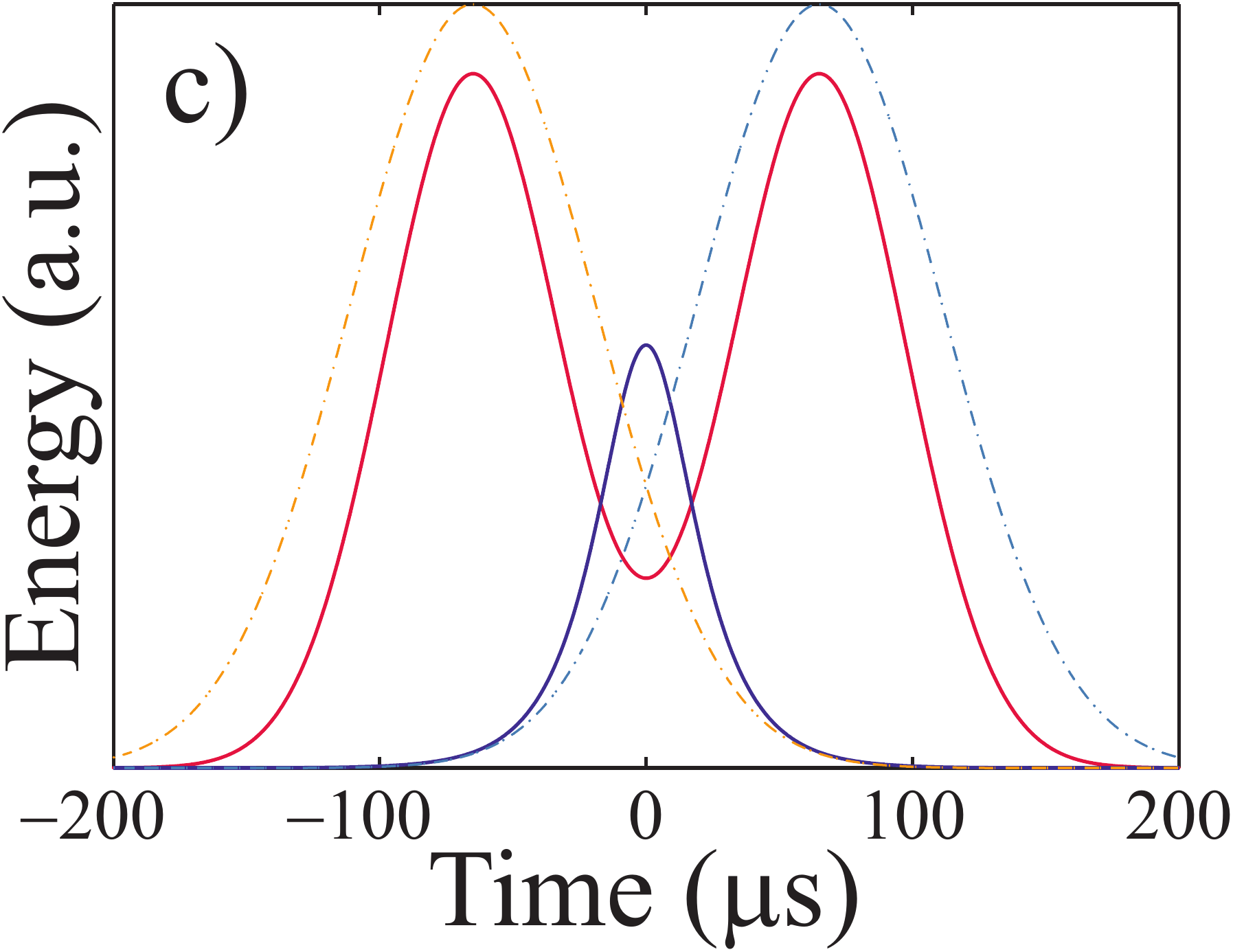}
	\label{fig:ac_timedep_sf13}
\caption[Time dependence of the adiabatic criterion]{\textbf{Time dependence of the adiabatic criterion} The couplings (left side of (\ref{eq:adiabatic_criterion_time_dep}), blue) and the energy splitting (right side of (\ref{eq:adiabatic_criterion_time_dep}), red) are shown for a fixed pulse length of \unit{100}{$\mu$s} and (a) a delay time of \unit{30}{$\mu$s}, (b) a delay time of \unit{80}{$\mu$s} and (c) a delay time of \unit{130}{$\mu$s}. The dotted lines represent the pulses.}
\label{fig:ac_timedep}
\end{figure}

In the following, we briefly review the basics of STIRAP in a 3-level $\Lambda$-scheme as shown in  \Cref{fig:3lvl_pulses}.
The Hamiltonian of the 3-level system coupled by two light fields in the interaction picture using the rotating wave approximation is given by \cite{bergmann_coherent_1998}: 
\begin{equation}
\mathcal{H}=\frac{\hbar}{2}
\left(\begin{array}{@{}ccc@{}}
-2\Delta_p & \Omega_p & 0\\ 
\Omega_p & 0 & \Omega_s\\ 
0 & \Omega_s & -2\Delta_s
\end{array}\right),
\label{eq:Hamiltonian}
\end{equation}
where the $\Omega_i$'s are the Rabi frequencies and the $\Delta_i$'s are the detunings of the so-called pump and Stokes laser beams with respect to the one-photon resonances. In the case of two-photon resonance $\Delta_p=\Delta_s=\Delta$, the eigenfrequencies of the system are given by
\begin{eqnarray}\label{eq:eigenvalue}
\omega_0&=&0,\nonumber\\
\omega_{+}&=&\frac{1}{2}\left(\Delta+\sqrt{\Delta^2+\Omega_p^2+\Omega_s^2}\right),\\
\omega_{-}&=&\frac{1}{2}\left(\Delta-\sqrt{\Delta^2+\Omega_p^2+\Omega_s^2}\right).\nonumber
\end{eqnarray}
For large detuning $\Delta\gg\Omega_i$, the corresponding eigenvectors (dressed states) become:
\begin{eqnarray}\label{eq:eigenstates}
\ket{a^0}&=&\cos\Theta\ket{1}-\sin\Theta\ket{3}\nonumber\\
\ket{a^+}&=&\ket{2}\\
\ket{a^-}&=&\sin\Theta\ket{1}+\cos\Theta\ket{3},\nonumber
\end{eqnarray}
where the so-called mixing angle $\Theta$ has been introduced. It is related to the Rabi frequencies of the coupling lasers by:
\begin{equation}
\tan\Theta=\frac{\Omega_p}{\Omega_s},
\label{eq:mixing_angle}
\end{equation}
The basic principle of the adiabatic transfer can be understood from the eigenstate equations (\ref{eq:eigenstates}). At the beginning of the sequence, only the Stokes laser field interacts with the atom and the adiabatic state $\ket{a^0}$ is aligned parallel to the initially populated electronic ground state $\ket{1}$. Due to the presence of the Stokes laser field, the initially degenerate energies of the system, $\omega_{-}$ and $\omega_{0}$, are split by the ac Stark shift. As long as this energy splitting is large compared to the coupling between the eigenstates of the system, no transition to other states will occur and the system stays in its instantaneous eigenstate, as stated by the adiabatic theorem \cite{born_beweis_1928}. By ramping the strength of the relative couplings between the three states such that only the pump laser induces a significant coupling at the end of the sequence (see \cref{fig:3lvl_pulses}b)), we can change the mixing angle $\Theta$ from $0$ to $\pi/2$. in doing so, we rotate the dressed state basis with respect to the bare state basis by $\unit{90}{^\circ}$, which means rotating $\ket{a^0}$ around $\ket{2}$ from $\ket{1}$ to $-\ket{3}$. If adiabaticity is maintained during the process, the population will stay in the eigenstate $\ket{a^0}$ and will follow the rotation, transferring it from the bare state $\ket{1}$ to state $\ket{3}$ without populating state $\ket{2}$. From  equations \ref{eq:eigenstates} we can see that the large detuning results in a symmetry of the eigenstates such that the dressed state $\ket{a^-}$ can be used as the initial state for population transfer using the so-called intuitive pulse order, which is sometimes called b-STIRAP in the literature. As mentioned above, the adiabatic criterion has to be fulfilled in the STIRAP sequence, i.e. the energy splitting must be larger than the couplings between the states \cite{gaubatz_population_1990, moller_efficient_2007}:
\begin{equation}
\bra{a^0}\frac{d}{dt}\ket{a^{\pm}}\ll\left|\omega^{\pm}-\omega^{0}\right|
\label{eq:adiabatic_criterion}
\end{equation}
The left side of the equation can be evaluated and it reads for the two states $\ket{a^{\pm}}$:
\begin{eqnarray}
\bra{a^0}\frac{d}{dt}\ket{a^{+}}&=0,\\
\bra{a^0}\frac{d}{dt}\ket{a^{-}}&=-\dot{\Theta}=\frac{\dot{\Omega}_p\Omega_s-\Omega_p\dot{\Omega}_s}{\Omega_{p}^{2}+\Omega_{s}^{2}}
\end{eqnarray}
Here we see that transitions to state $\ket{2}=\ket{a^+}$ are not allowed due to the large detuning. We now insert the second equation and the eigenvalues (\ref{eq:eigenvalue}) of the Hamiltonian in (\ref{eq:adiabatic_criterion}) and get a time dependent adiabatic criterion:
\begin{equation}
\frac{\dot{\Omega}_p\Omega_s-\Omega_p\dot{\Omega}_s}{\Omega_{p}^{2}+\Omega_{s}^{2}}\ll\frac{1}{2}\left(\Delta-\sqrt{\Delta^2+\Omega_p^2+\Omega_s^2}\right)
\label{eq:adiabatic_criterion_time_dep}
\end{equation}
We plotted both sides of the equation for a pulse length of \unit{100}{$\mu$s} and three different delay times of \unit{30}{$\mu$s}, \unit{80}{$\mu$s} and \unit{130}{$\mu$s} in \cref{fig:ac_timedep}.
For short delay times (\cref{fig:ac_timedep}a)), the adiabatic criterion is not fulfilled at the beginning and the end of the pulse sequence. During this part of the sequence transitions between the adiabatic states $\ket{a^0}$ and $\ket{a^+}$ may occur, leading to non-adiabatic transfer that depends on the relative populations in the two involved adiabatic states. For long delay times (\cref{fig:ac_timedep}c)), the adiabatic criterion is not fulfilled in between the two pulses. 

\section{\label{sec:level3} Experimental Setup}
Details of the experimental setup have been described before \cite{hemmerling_single_2011,hemmerling_novel_2012}. We use the $\ket{F=3, m_F=3}=\ket{\!\!\downarrow}=\ket{1}$, $\ket{F=2, m_F=2}=\ket{\!\!\uparrow}=\ket{3}$ states of the ${}^2$S$_{1/2}$-ground state of a \Mg{25} ion as our qubit. The states are separated in energy by the hyperfine splitting of \unit{1.789}{GHz}. The frequency quadrupled output of a fiber laser is used to create the laser beams at a wavelength of 280~nm for Doppler cooling, Raman sideband cooling and for coherent manipulation. The first order sideband of an electo-optic modulator (EOM) is resonant with the ${}^2$S$_{1/2}$ $\rightarrow$ ${}^2$P$_{1/2}$ transition for Doppler cooling and state discrimination. The \unit{9.2}{GHz} red detuned optical carrier is used with an additional acousto-optical modulator (AOM) setup to create the Raman laser beams that couple the hyperfine qubit states (see \cref{fig:3lvl_pulses}). Additionally, a radio frequency can be applied to couple the qubit states without being influenced by or changing the motional state. A sequence of consecutive Raman red sideband and repump pulses is used for ground state cooling the axial vibrational mode of the ion \cite{wan_efficient_2015}.\\
We implemented the STIRAP sequence in our setup using a pulse sequencer based on a field programmable gate array (FPGA) \cite{pham_general-purpose_2005, schindler_frequency_2008} that controls direct digital synthesizer (DDS) boards. We used the built-in power sweep function of the sequencer to shape the amplitude of the two laser beams needed for the STIRAP sequence. It is implemented by applying a voltage resembling the shape of the desired pulse to a voltage controlled gain amplifier which modulates the rf signal generated by the DDS chip. This radio frequency signal is subsequently amplified and fed into an AOM that imprints the time dependence of the radio frequency amplitude onto the laser intensity. 
The peak resonant Rabi frequency for each beam is on the order of a few 10 megahertz, resulting in carrier Raman Rabi frequencies of around 100~kHz. The experimental data is typically averaged over 250 repetitions of an identical experiment with the same initial conditions.


\section{\label{sec:level4} STIRAP simulation}
Numerical simulations were carried out based on the density matrix formalism to determine the parameter regime for efficient population transfer using the STIRAP process. We integrated the master equation numerically and derived the time dependence of the atomic state populations. In general, the master equation can be expressed as:\\
\begin{equation}
\frac{d\rho}{dt}=\mathcal{L}\rho.
\end{equation}
Here, $\mathcal{L}$ is the Liouvillian operator. Our qubit states $\ket{1}$ and $\ket{3}$ are magnetic sub-states of the hyperfine splitted ground state of the \Mg{25} ion and spontaneous emission from these long-lived states is neglected in the simulation. The detuning of the lasers with respect to the excited state $\ket{2}$ is \unit{9.2}{GHz} and the off-resonant scattering rates from coupling to all possible excited states are on the order of \unit{1.1}{ms$^{-1}$} and \unit{4.4}{ms$^{-1}$} for the ground states $\ket{1}$ and $\ket{3}$, respectively \cite{gebert_damage-free_2014}. This off-resonant scattering limits the coherence between the lasers and the atom and reduces the detected signal. However, this effect is small and since the consideration of off-resonant scattering increases the complexity of the simulation excessively, we neglected this effect in the simulations. In this case the time evolution of the system can be described by a Hamiltonian $\mathcal{H}$ and the Liouvillian acts on the density matrix as follows:
\begin{equation}
\mathcal{L}\rho=-i\left[\mathcal{H},\rho\right]=-i\left(\mathcal{H}\rho-\rho\mathcal{H}\right)
\label{eqn:Liouvillian}
\end{equation}
The quantized motion of the ion in the trap is included in the simulations as the tensor product of the electronic state \ket{e} and the harmonic trap states \ket{n} for the state of the ion: $\ket{\psi}=\ket{e}\otimes\ket{n}$. Up to 16 motional levels have been considered.
This way, we are able to simulate carrier as well as sideband transitions. The time dependence of the STIRAP process is incorporated by time-dependent Rabi frequencies in the Hamiltonian, where for the simulation a Gaussian pulse shape was assumed. The parameters of the pulses are defined as:
\begin{eqnarray}
\Omega_{i}(t)&=&\Omega_{i,\mathrm{max}}\cdot\exp\left(-\frac{\left(t-t_i\right)^2}{2t_\mathrm{width}^2}\right)
\end{eqnarray}
where we define $t_\mathrm{pulse}=2\sqrt{2 \ln(2)}\cdot t_\mathrm{width}$ as the pulse length, $t_{i}$ as the centers, and $\Omega_{i,\mathrm{max}}$ as the maximum Rabi frequencies of the two pulses $i\in\{p,s\}$. Additionally, we denote the delay between the pulses as $t_\mathrm{delay}=t_s-t_p$ (see \cref{fig:3lvl_pulses}).
All simulations presented in the following were performed using the quantum optics toolbox \cite{tan_computational_1999} in the MATLAB programming language \cite{matlab_version_2013}. Since the Hamiltonian of the system is time dependent, the solver "\textbf{solvemc}" was used. It performs a direct integration of the master equation to calculate the density matrix $\rho$ for consecutive times.

In order to compare the simulated with the experimental results we fitted a Gaussian function to the measured pulses. Due to technical imperfections the measured pulse length of the pump field is around 12~\% shorter than the pulse length of the Stokes field. Therefore we derive an effective pulse length (mean of the Gaussian full width at half maximum, FWHM) from the fit and measure the effective delay time of the two pulses. These values, which were ranging from a few to two hundred microseconds, were used for the laser pulse parameters in the simulations.

\section{\label{sec:level5} Results}
\subsection{\label{sec:level5a} Pulse length and pulse delay dependence}
Density matrix simulations were performed to investigate and optimize the population transfer efficiency. First, the influence of the delay of the two laser pulses for a fixed pulse length of \unit{120}{$\mu$s} was studied. The ion is initialized in state \ket{1} and the motional ground state.
\begin{figure}[tbp]
	\centering
\includegraphics[width=0.46\textwidth]{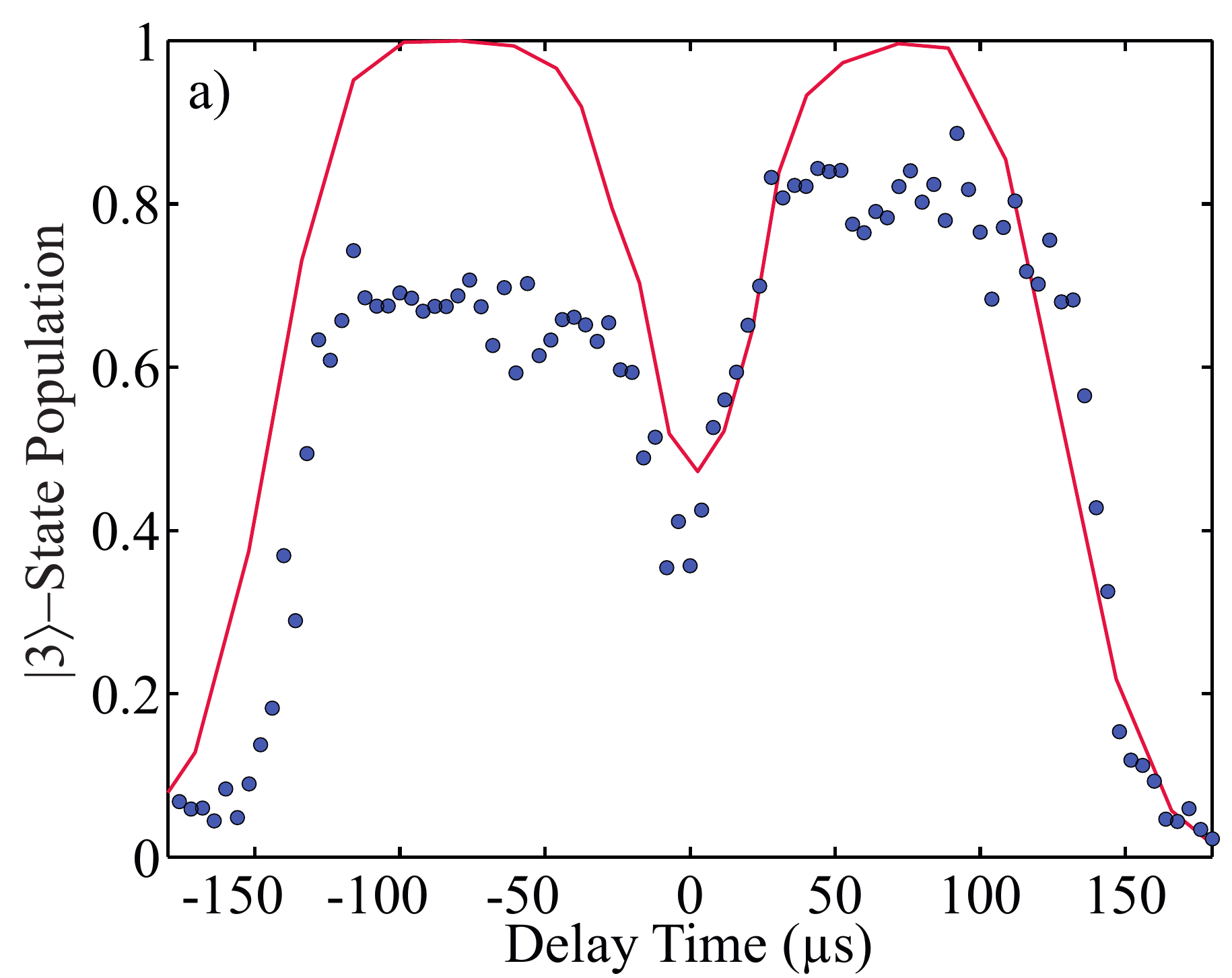}
	\label{fig:comp_sim_meas_delay}
\includegraphics[width=0.46\textwidth]{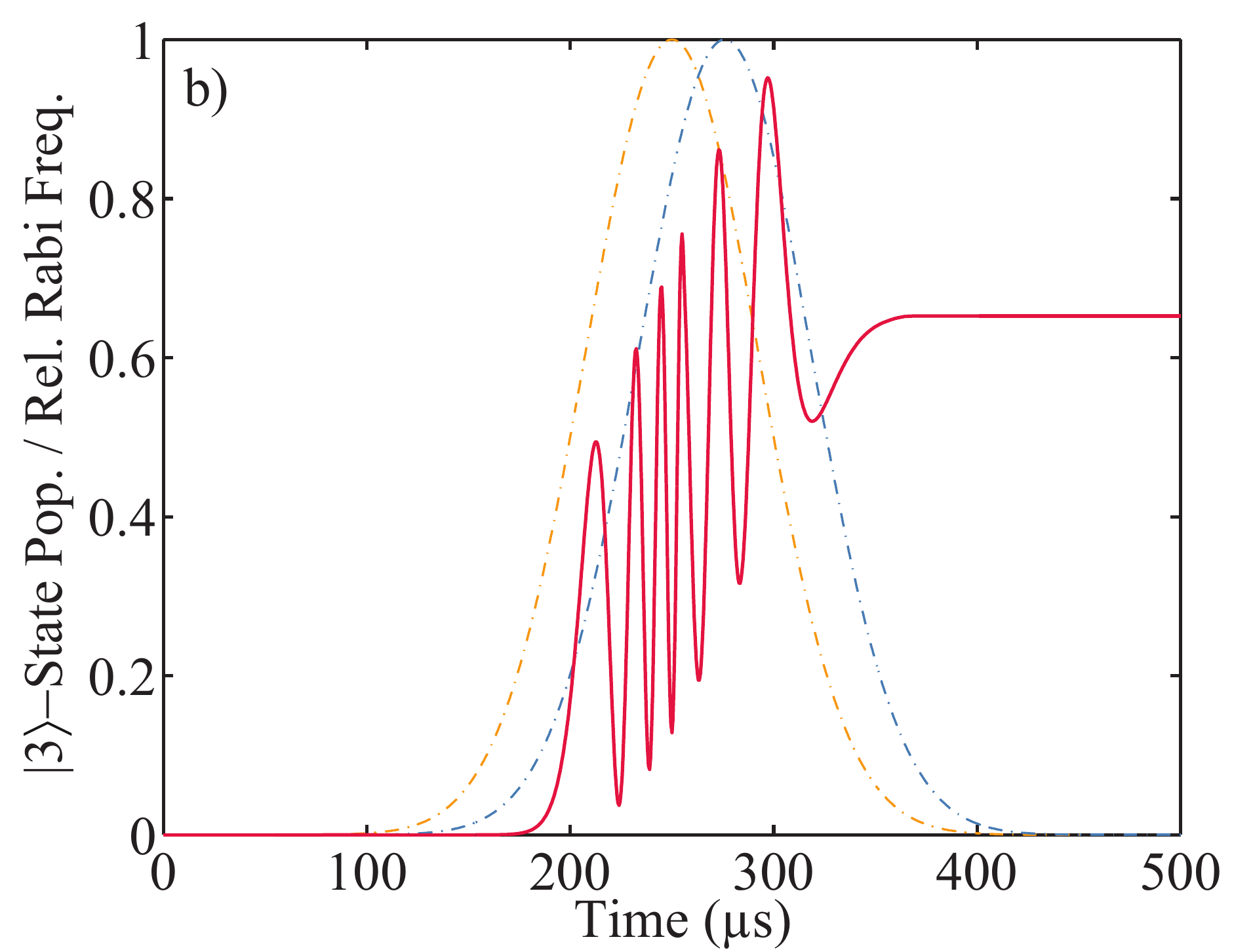}
	\label{fig:sim_trans_delay26}
\includegraphics[width=0.46\textwidth]{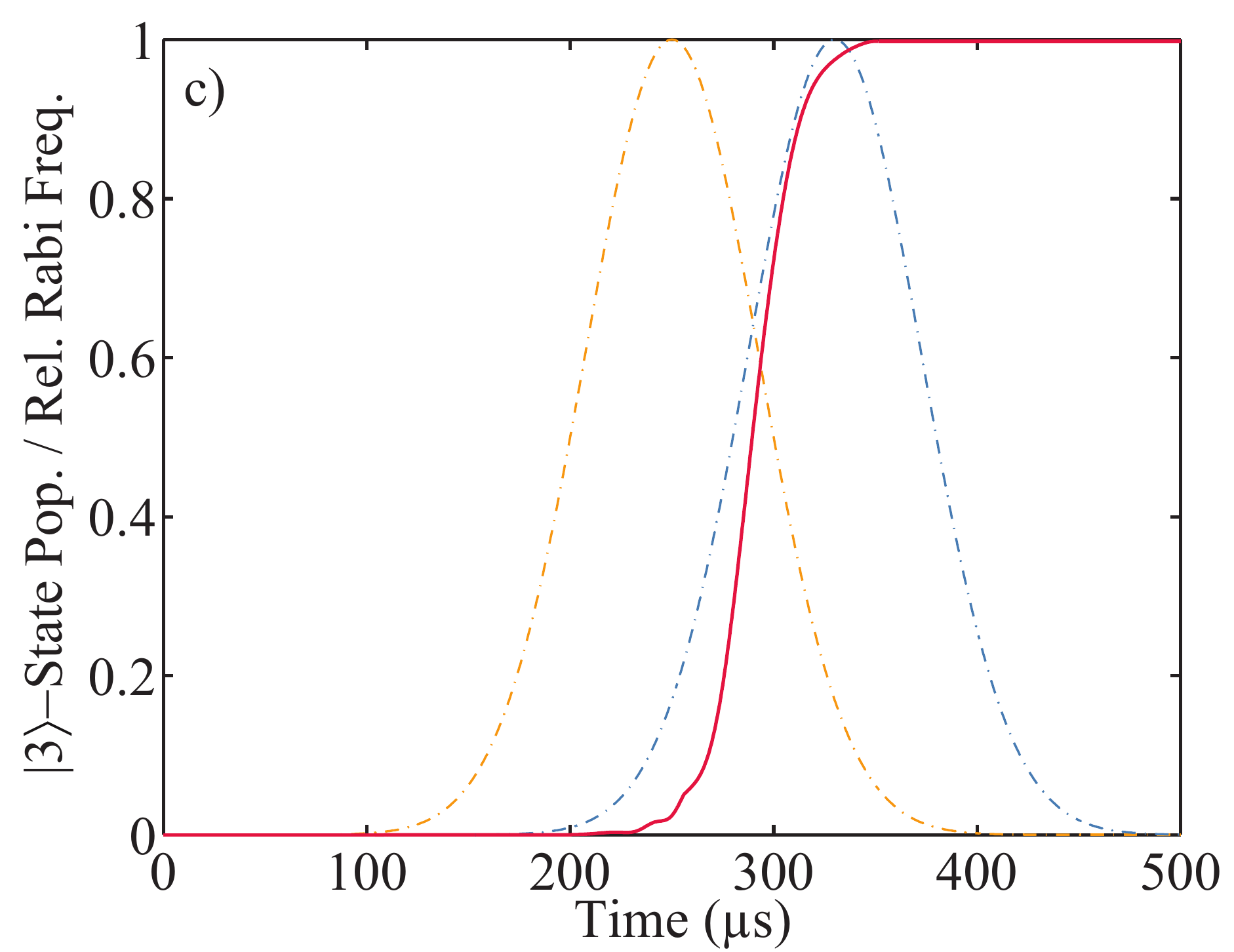}
	\label{fig:sim_trans_delay80}
\includegraphics[width=0.46\textwidth]{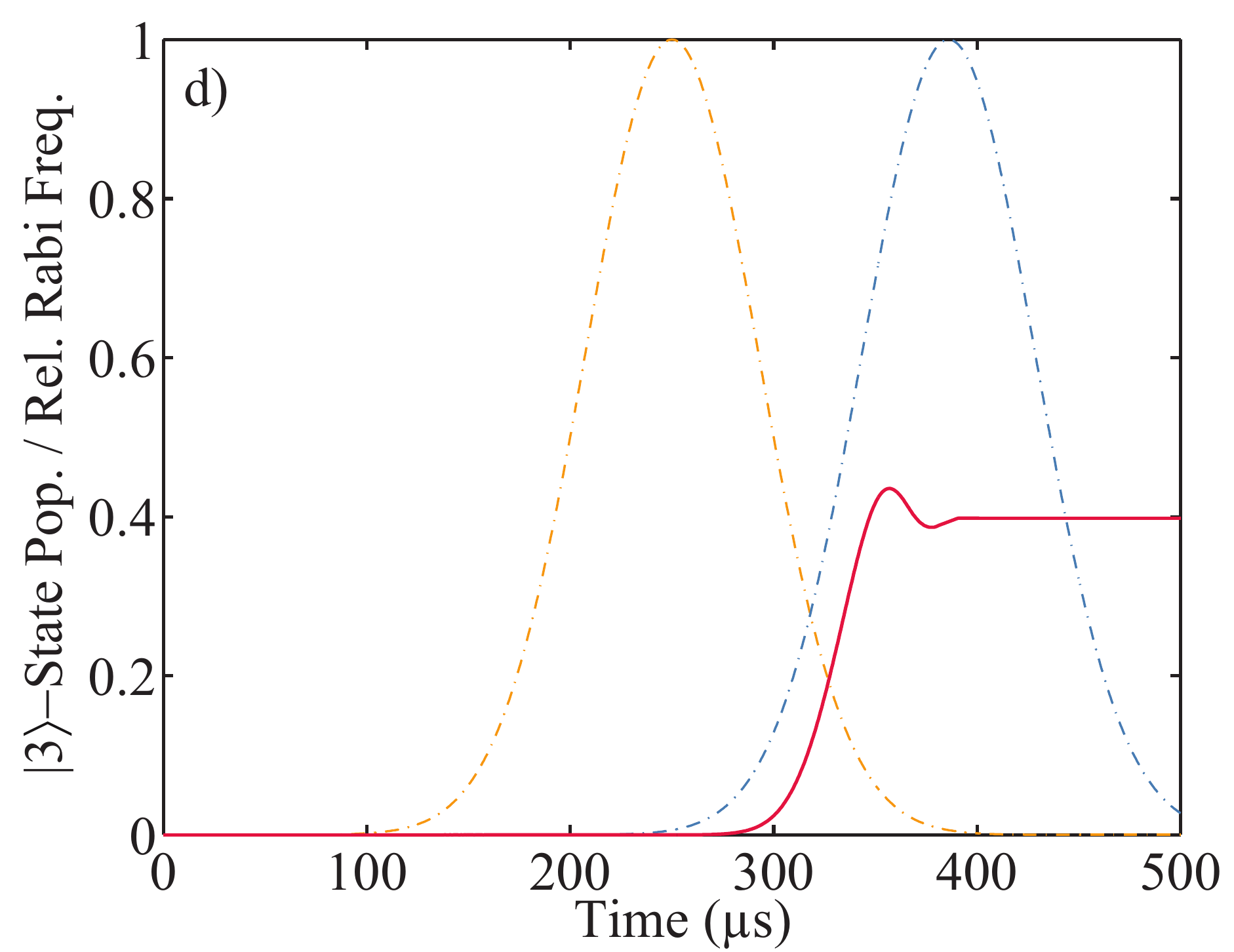}
	\label{fig:sim_trans_delay136}
\caption[Simulated and measured delay time dependence of the STIRAP transfer.]{\textbf{Simulated and measured delay time dependence of the STIRAP transfer.} a) Comparison of the simulated and measured delay time dependence of the STIRAP transfer on a carrier transition for an ion in the motional ground state. Positive delay times correspond to the counter-intuitive pulse sequence. For the simulated data (red) a moving average was used to account for experimental intensity fluctuations. The simulated time dependence of the STIRAP transfer for different delay times is shown in b) for \unit{26}{$\mu$s}, c) for \unit{80}{$\mu$s}, and d) for \unit{136}{$\mu$s}. The dotted lines represent the pulses.}
\label{fig:delay_time_dep}
\end{figure}
In \cref{fig:delay_time_dep}a) we compare the measured and simulated transfer efficiency of the STIRAP sequence for carrier transitions. To allow a direct comparison between simulation and experiment, a moving average is applied to the simulated data to account for small fluctuations of experimental parameters (see below).
In the figure we see that both pulse orders (positive delay times correspond to the counter-intuitive pulse order) are able to transfer population from the initial to the final state. This behavior is a result of the large detuning of the light fields from resonance. 
The reduced transfer efficiency seen in the experiment compared to the simulation result is explained by off-resonant excitation to different magnetic sub-states, indicated in \cref{fig:3lvl_pulses}a) which is not considered in the simulations. The relative populations of the two qubit states during the STIRAP sequence, together with the relative strength of the involved couplings of these states to the auxiliary magnetic sub-states \cite{gebert_damage-free_2014}, explain the asymmetry with respect to the pulse order seen in the experiment. We therefore use the counter-intuitive pulse sequence from now on.

Using the numerical simulations we investigate the different transfer regimes. As can be seen in \cref{fig:delay_time_dep}b), for a delay time of $\unit{26}{\mu s}$, the transfer process consists of two contributions, one oscillating and one adiabatic part. For the chosen pulse length of $\unit{120}{\mu s}$, the adiabatic criterion is not fulfilled at the beginning and end of the sequence [see \cref{fig:ac_timedep}a)] and the time evolution can be understood as a combination of dynamically varying Rabi oscillations, which is the transfer mechanism for the case of overlapping pulses (zero delay time), and an adiabatic part of the population transfer. In this situation the final transfer efficiency strongly depends on the timing and exact Rabi frequency of the pulses. Fluctuations of experimental parameters result in an averaged transfer efficiency. 
Full adiabaticity is achieved for a delay time of for example $\unit{80}{\mu s}$, shown in \cref{fig:delay_time_dep}c) [see also \cref{fig:ac_timedep}b)]. For this case efficient adiabatic transfer is possible, whereas for an even longer delay time of $\unit{136}{\mu s}$ the overlap of the pulses is too small, leading to transitions between the different adiabatic states [\cref{fig:ac_timedep}c) and \cref{fig:delay_time_dep}d)], so that the transfer is incomplete.

The optimal parameter regime for STIRAP was investigated by comparing the experimental and simulated population transfer with respect to the pulse length for different delays between the pulses. For this we introduce a scaling factor $s$ for the delay time which is related to the pulse length by $t_\mathrm{delay}=s \cdot t_\mathrm{pulse}$. If the pulse length is chosen long enough, we are in the Rabi oscillation regime for $s<0.6$ and in the adiabatic (STIRAP) regime for $s>0.6$. 

\begin{figure}[tbp]
	\centering
\includegraphics[width=0.46\textwidth]{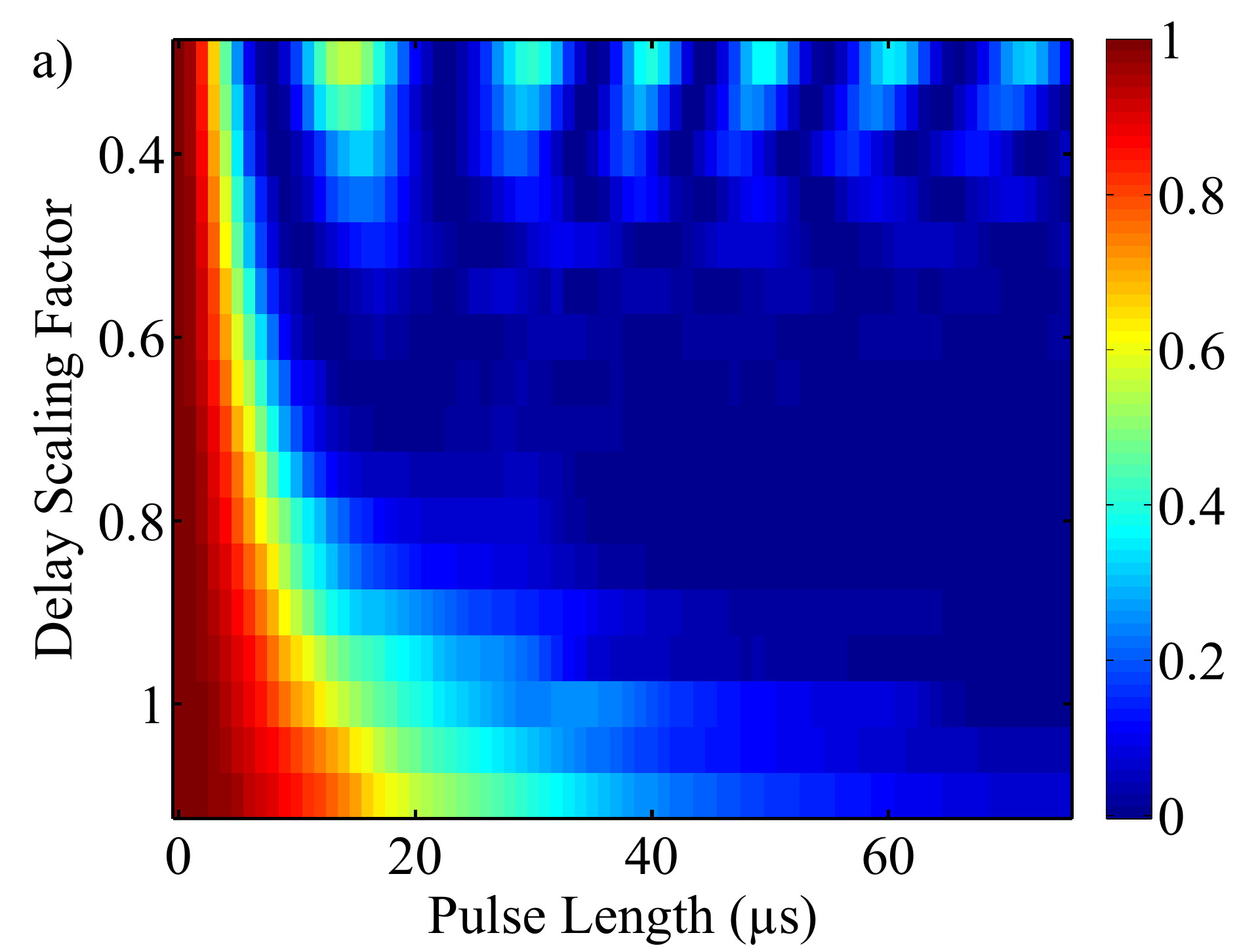}
	\label{fig:pop_vs_delay_and_pulselength}
\includegraphics[width=0.46\textwidth]{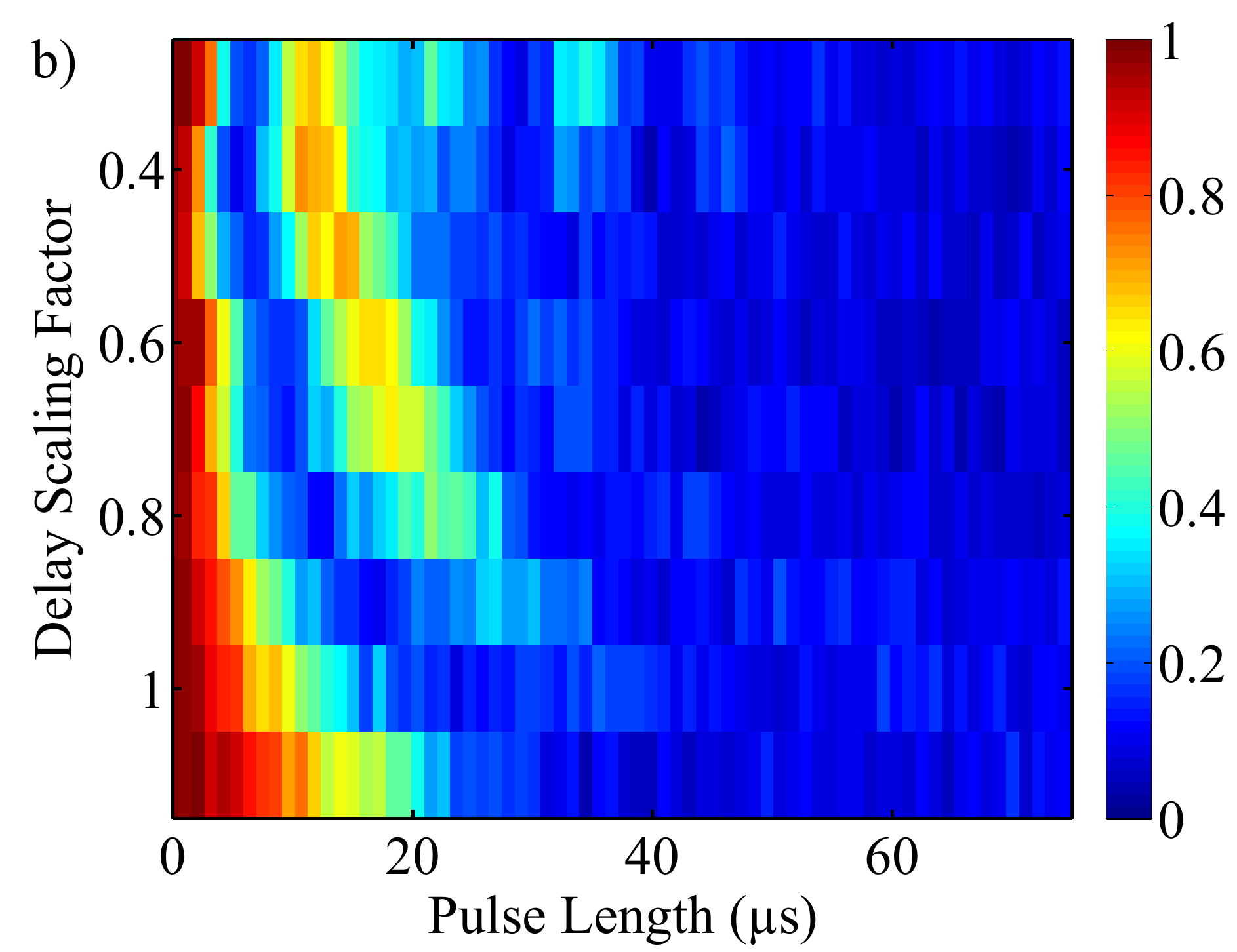}
	\label{fig:pop_vs_delay_and_pulselength_meas}
\caption[Transfer efficiency for different delay scaling factors and pulse lengths for carrier transitions.]{\textbf{Transfer efficiency for different delay scaling factors and pulse lengths for carrier transitions.} a) Simulation and b) experimental results of the STIRAP transfer, where red corresponds to no and blue to complete transfer. For the simulations and the measurements the ion was initialized in the ground state of motion in \ket{3} and the pulse length was scanned for different delay scaling factors $s$.}
\label{fig:2D_plots_car}
\end{figure}

In \cref{fig:2D_plots_car} the simulated transfer efficiency for carrier transitions with the ion initialized in the motional ground state is compared with the experimental result. 
Experiment and simulation show the same behavior as in \cref{fig:delay_time_dep}, i.e. the transfer depends on the exact pulse length for small delay scaling factors. The transfer is slightly faster in the experiment than in the simulations which may be due to slight deviations of the used parameters as well as an imperfect pulse shape of the STIRAP pulses. Additionally, deviations from the ideal pulse shape may lead to the enhanced oscillations in the experiment, since in the experiment the pulses envelope is less smooth. Despite these differences, the experimental data is in qualitative agreement with the simulations and we can identify large regions of efficient transfer for the constant coupling strength realized in this particular setting. The results also allow to extrapolate which STIRAP parameters should be used for the case of a fluctuating coupling strength, since the required pulse length scales with the inverse coupling strength. For a ``band'' of coupling strengths, the optimum STIRAP transfer parameters are dictated by the corresponding graph for the smallest coupling strength. Efficient transfer is then inherently accomplished for the larger coupling strengths.

\begin{figure}[tbp]
	\centering
\includegraphics[width=0.46\textwidth]{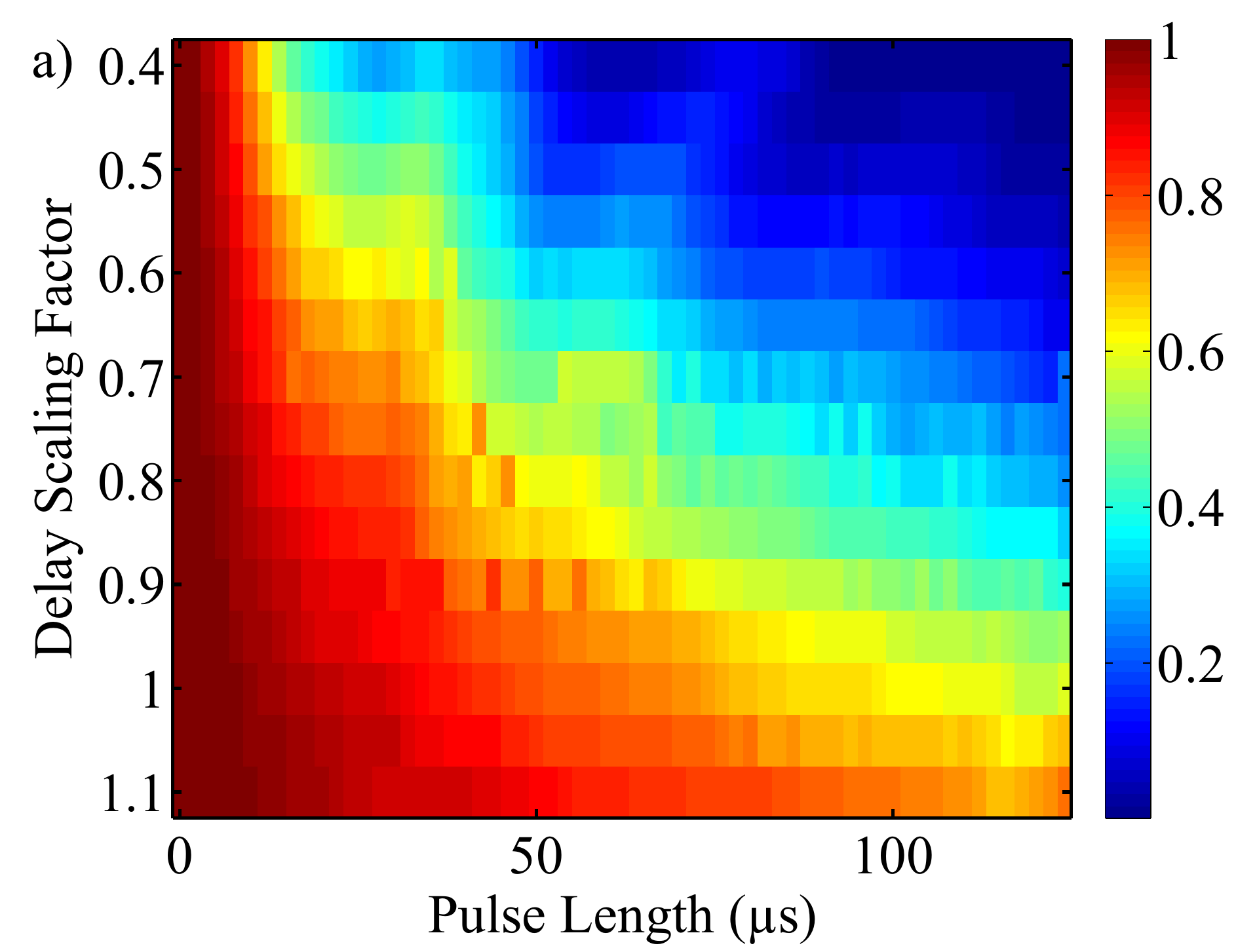}
	\label{fig:pop_vs_delay_pulselength_rsb}
\includegraphics[width=0.46\textwidth]{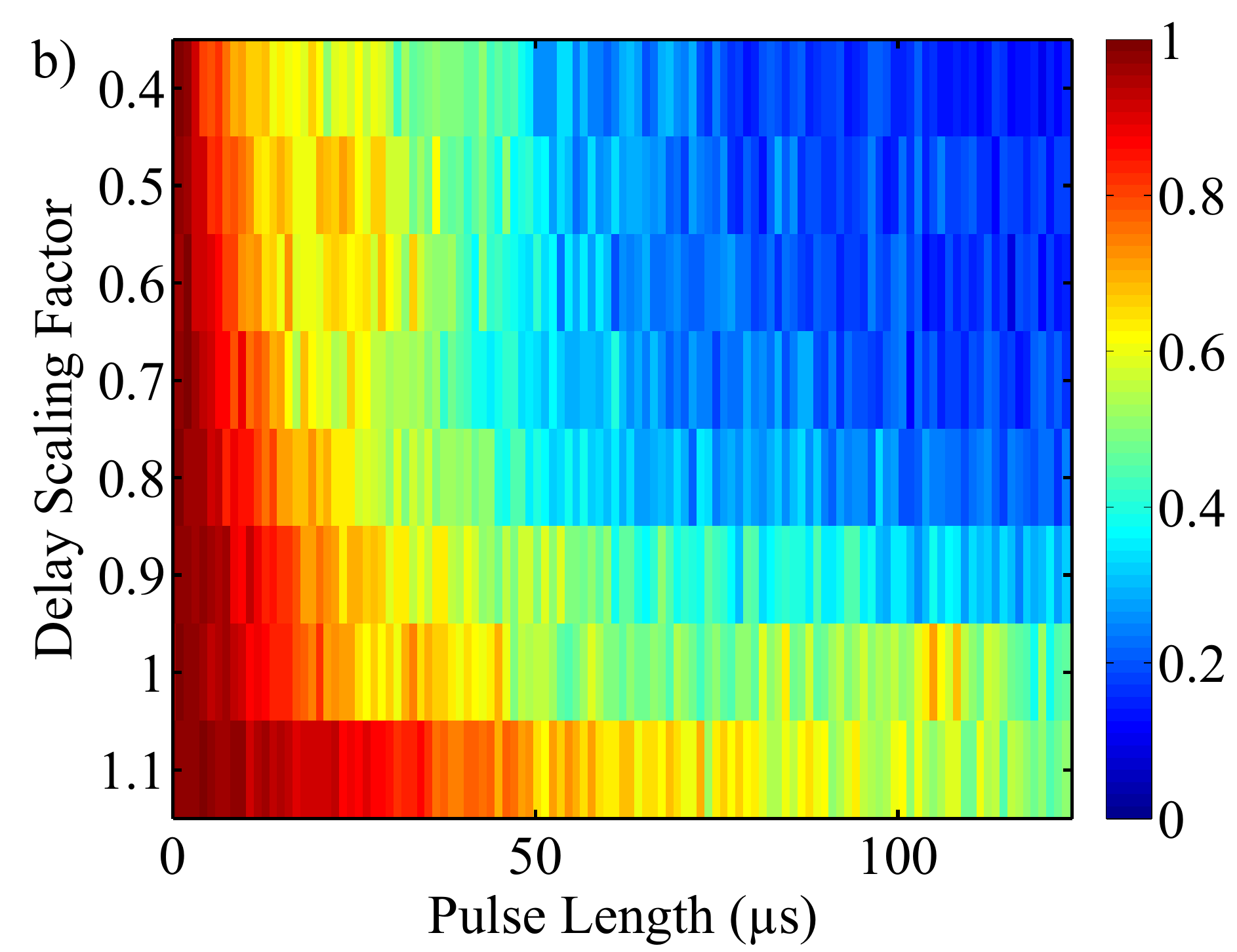}
	\label{fig:pop_vs_delay_pulselength_rsb_meas}
\caption[Transfer efficiency for different delay scaling factors and pulse length for blue sideband transitions.]{\textbf{Transfer efficiency for different delay scaling factors and pulse length for blue sideband transitions.} a) Simulation and b) experimental results of the STIRAP transfer, where red corresponds to no and blue to complete transfer. For the simulations and the measurements the ion was initialized in the ground state of motion and the pulse length was scanned for different delay scaling factors $s$.}
\label{fig:2D_plots_bsb}
\end{figure}

This scaling behavior is illustrated in \cref{fig:2D_plots_bsb}, where the transfer efficiency is shown for a blue sideband transitions with the ion initialized in the motional ground state. The sideband transition is weaker by the Lamb-Dicke factor of approximately $\eta\approx 0.3$.
Therefore, a longer pulse length is required to fulfill the adiabatic criterion and to achieve efficient transfer. As can be seen in \cref{fig:2D_plots_bsb}, for this transition we can also identify a parameter range, where we have efficient transfer for our given laser power and detuning. For the carrier transition we achieve the best transfer for a delay scaling factor of around $\textit{s}=0.7$ and a pulse length  $>\unit{50}{\mu s}$, whereas for sideband transitions smaller scaling factors and longer pulse lengths are necessary. For both transitions off-resonant excitation to the excited $^2P_{3/2}$-state is limiting the final transfer efficiency. The longer pulses required for sideband transitions limit the transfer efficiency to \unit{85}{\%} for a pulse duration of $\unit{>70}{\mu s}$ and a delay scaling factor of $\leq0.7$. 

\subsection{\label{sec:level5b} Motional Dependence of the Transfer Efficiency}
\begin{figure}[tbp]
	\centering
\includegraphics[width=0.46\textwidth]{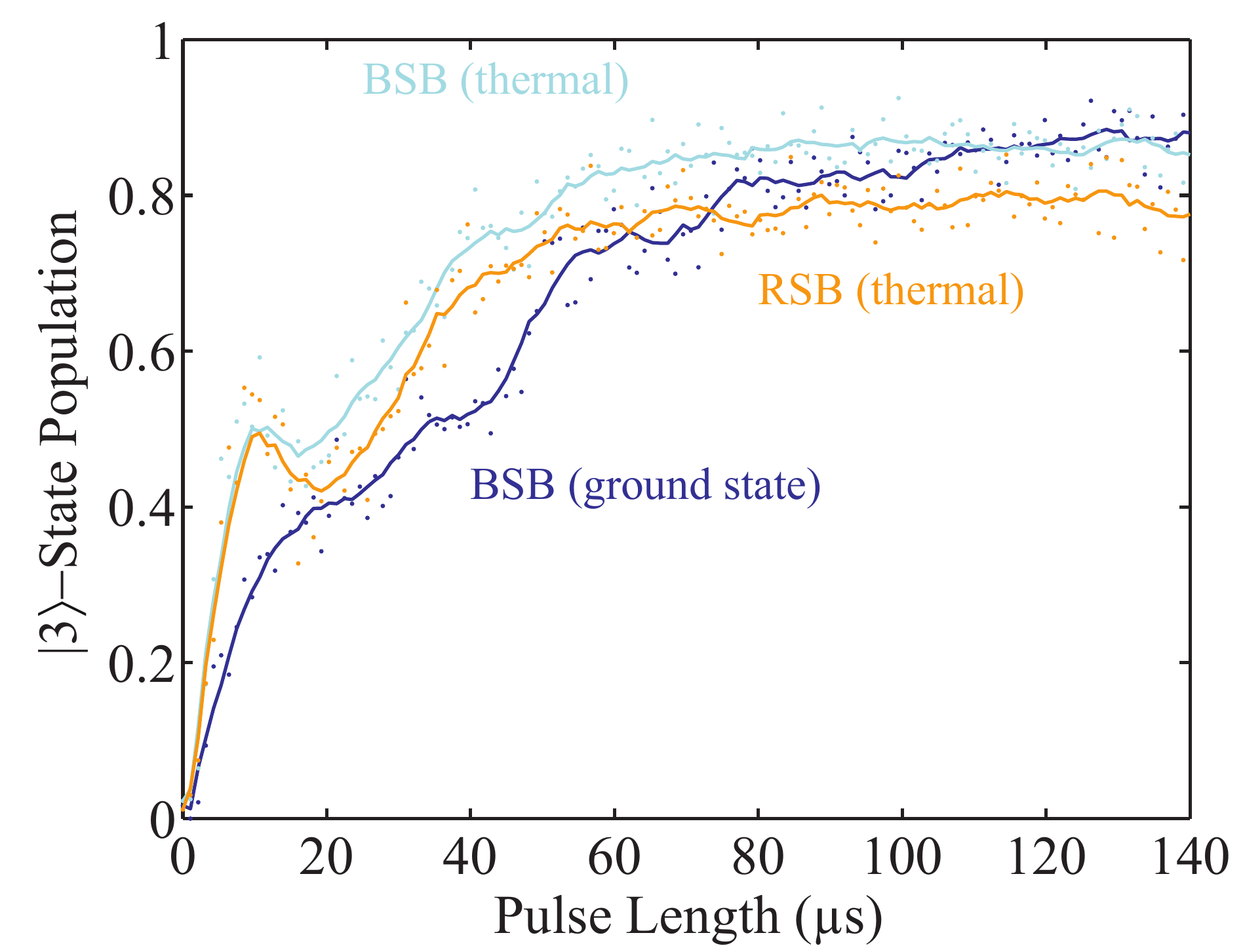}
	\caption[STIRAP transfer efficiency as a function of pulse length for different initial motional states.]{\textbf{STIRAP transfer efficiency as a function of pulse length for different initial motional states.} The ion was initialized in the motional ground state (dark blue) and a thermal state (light blue) for a blue sideband transition. Additionally the transfer is displayed for a red sideband transition, where the ion was initialized in a thermal state (orange). The delay scaling factor for all measurements was $s=0.5$. The lines are moving averages of the data and are guides to the eye. For clarity the error bars are omitted.}
	\label{fig:comp_thermal_with_ground_state}
\end{figure}

After determining appropriate parameters for the transfer we now investigate the motional dependence of the transfer efficiency. Using the simulations we investigated the dynamics of population transfer for the lowest 15 motional Fock states for carrier and sideband transitions.
\begin{figure}[tbp]
	\centering
\includegraphics[width=0.46\textwidth]{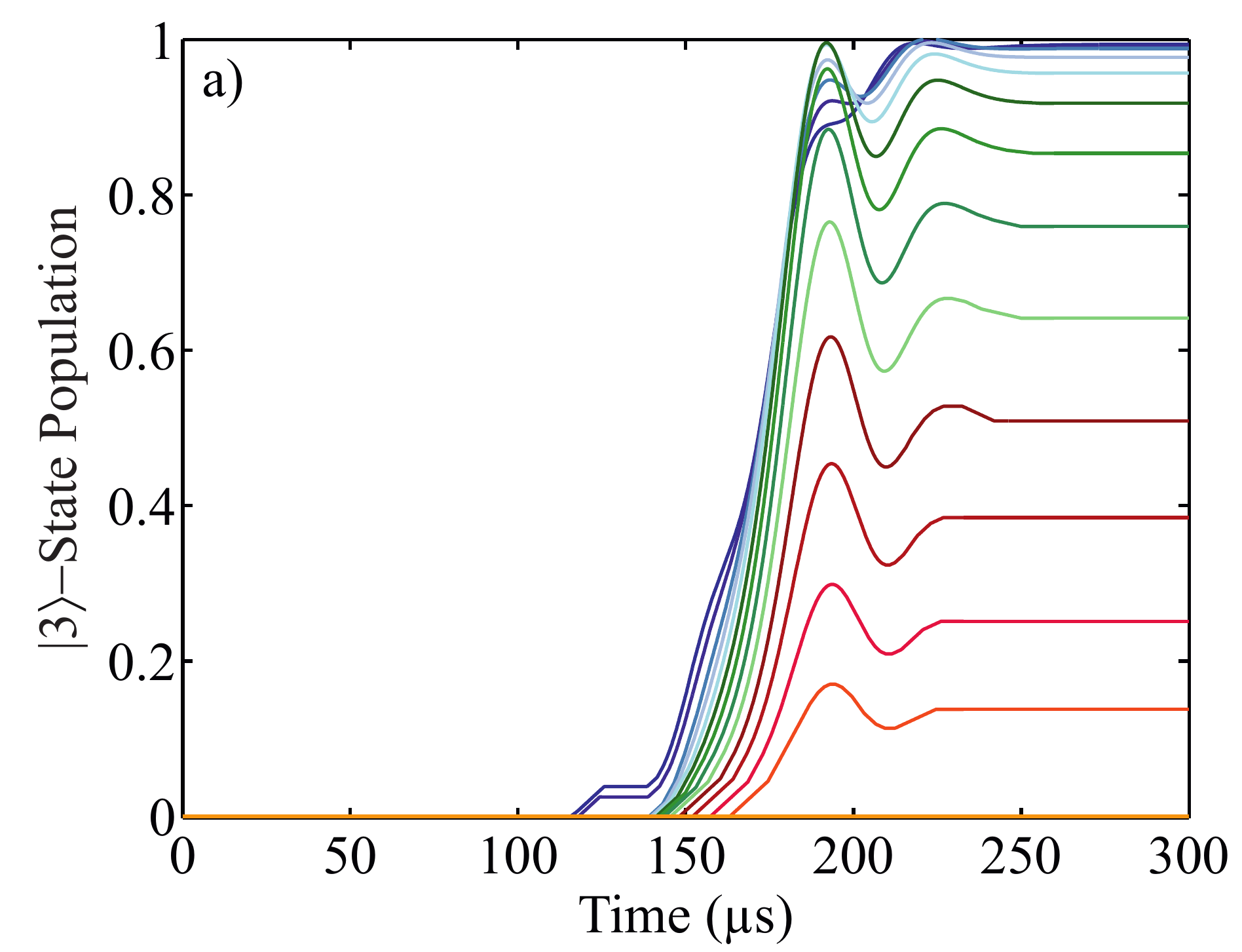}
	\label{fig:car_diff_n}
\includegraphics[width=0.46\textwidth]{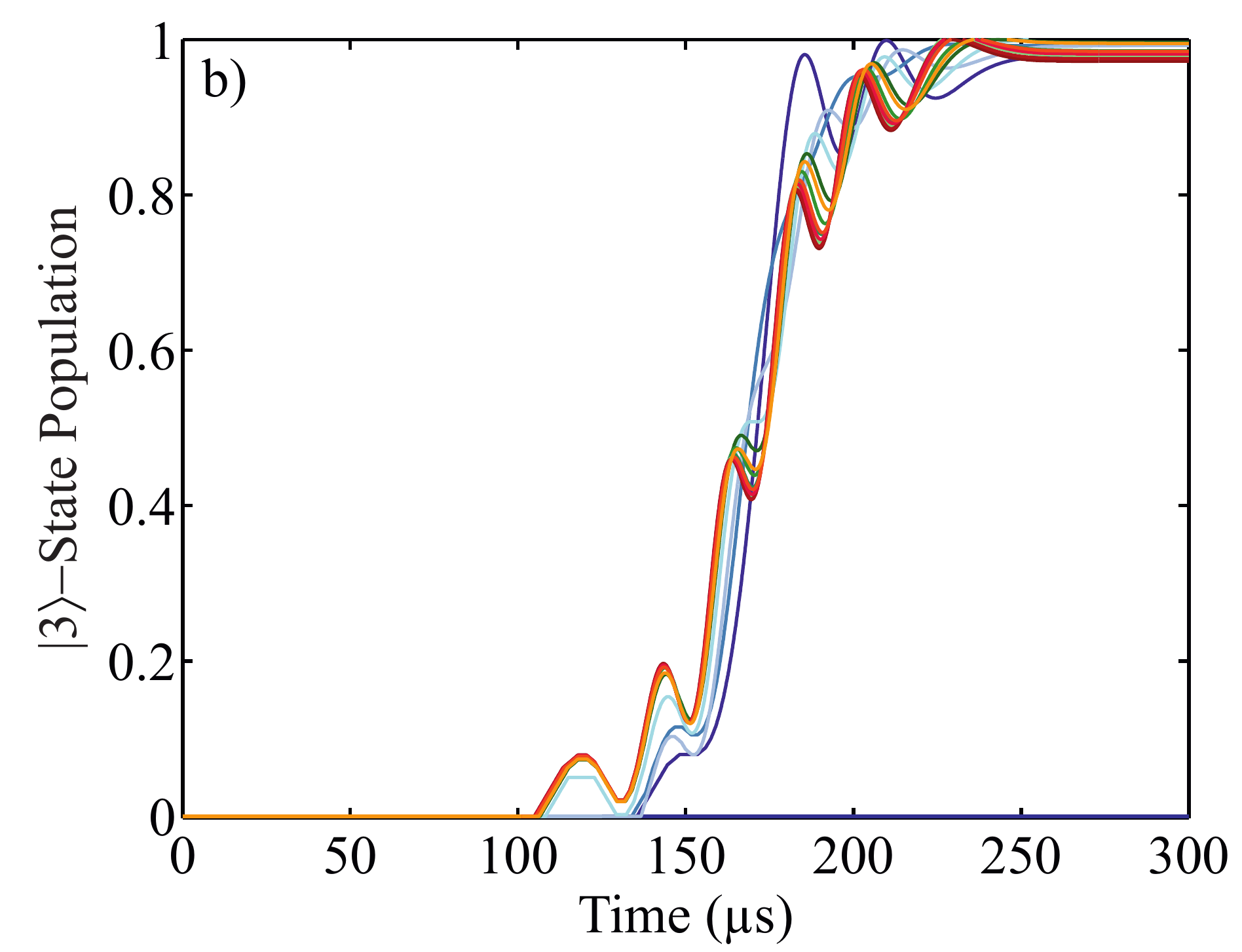}
	\label{fig:rsb_diff_n}
\caption[Simulated transfer dynamics of the population for the lowest 15 motional levels.]{\textbf{Simulated transfer dynamics of the population for the lowest 15 motional levels.} The population transfer as a function of time is displayed for a) the carrier (delay scaling factor $s=0.7$ and pulse length of $\unit{50}{\mu s}$) and b) the red sideband transition (delay scaling factor $s=0.4$ and pulse length of $\unit{100}{\mu }$). Fock states are used in the simulations, where dark blue corresponds to the motional ground state and the color gets brighter for higher motional levels and changes from blue to green, to red and finally to yellow for n=15.}
\label{fig:sim_time_evolution_diff_n}
\end{figure}
The results are shown in \cref{fig:sim_time_evolution_diff_n}, where we can see that for carrier transitions the transfer efficiency is reduced for higher motional levels. This is due to the decrease in carrier coupling strength associated with higher Fock state levels, scaling with the generalized Laguerre polynomial $L_n^0(\eta^2)$ \cite{wineland_experimental_1998}. In contrast, the coupling strength of blue sideband transitions initially increases with $n$ according to $\sqrt{n!/(n+1)!}L_n^1(\eta^2)$ and remains at a high level up to the largest investigated Fock state $n=15$. As a consequence, the transfer becomes more adiabatic for higher motional levels, since the state evolution speeds up while the pulse time remains constant. Experimentally, we do not probe each of the motional states individually, but rather a given distribution. Therefore, we experimentally verified the independence of the STIRAP transfer from the initial motional state  by investigating the pulse length dependence of the transfer efficiency for the ion initialized in the motional ground state and compare it to the efficiency for the ion initialized in a thermal state. Furthermore, the counter-intuitive pulse sequence allows us to further speed up the STIRAP sequence by considering only the part of laser interaction where the transfer takes place: The Stokes beam is switched on abruptly with the maximum intensity at the beginning of the sequence. While ramping down its intensity, the pump beam intensity is ramped up, both with a near-Gaussian shape. At maximum intensity of the pump beam, it is switched off rapidly. The effective transfer time for this sequence is given by $t_{trans}=t_{FWHM}+s\cdot t_{FWHM}$. As can be seen from \cref{fig:comp_thermal_with_ground_state}, the STIRAP population transfer rate for sideband transitions is smaller when the ion is initialized in the motional ground state. This is a consequence of the smaller Rabi frequency of sideband transitions with $n=0$ compared to $n>0$ also populated in a thermal state. For pulse lengths longer than $\unit{100}{\mu s}$, the two curves for the transfer on the blue sideband overlap with each other. This means that, for a sufficiently long pulse, the STIRAP transfer becomes independent of the motional state population of the ion.

Additionally, the transfer efficiency for a red sideband transition is shown in \cref{fig:comp_thermal_with_ground_state} for the ion initialized in a thermal state of motion. We assume that for a pulse length of $\unit{120}{\mu s}$ the transfer is complete. 
In contrast to a blue sideband, the red sideband leaves the $n=0$ motional population untouched. We extract a ground state population of $0.08\pm 0.01$ by subtracting the respective signals averaged over pulse lengths between $\unit{120}{\mu s}$ and $\unit{150}{\mu s}$. The associated thermal distribution corresponds to a temperature of $1.3\pm0.2$ times the Doppler cooling temperature for \Mg{25}. This is in agreement with the measured temperature after Doppler cooling of 1.2 times the Doppler temperature using the technique described in \cite{Poulsen_sideband_2011} as well as the estimated temperature from measured Rabi oscillation decay which resulted in $1.6\pm 0.5$ \cite{hemmerling_towards_2011}.


\subsection{\label{sec:level5c} Comparison of Coherent and Adiabatic Transfer}
After showing the feasibility of motional state-independent transfer we compare the transfer efficiency of the STIRAP process with that of a $\pi$ pulse for a Doppler cooled ion.

\begin{figure}[tbp]
	\centering
\includegraphics[width=0.46\textwidth]{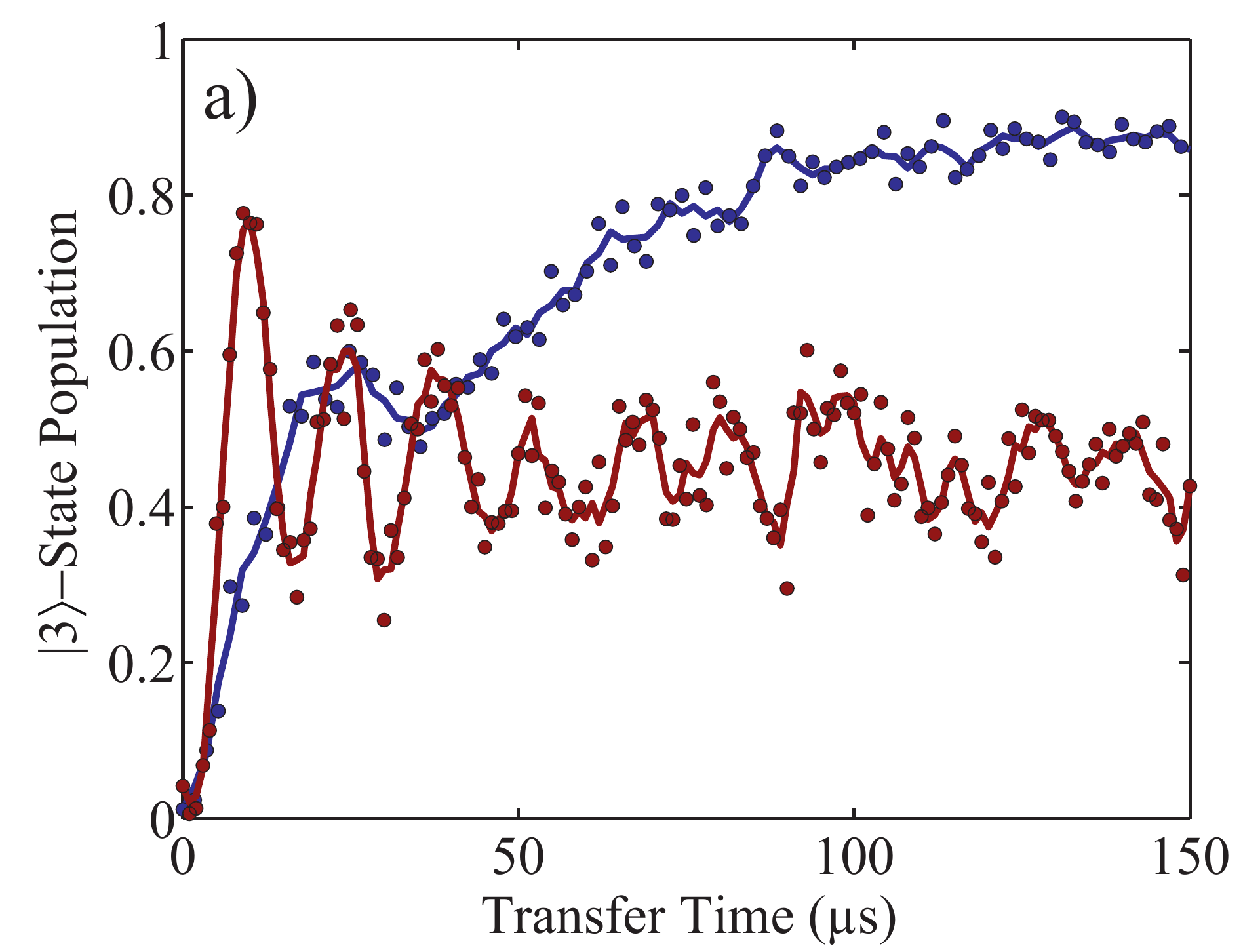}
	\label{fig:bsb_comp_stirap_rabi}
\includegraphics[width=0.46\textwidth]{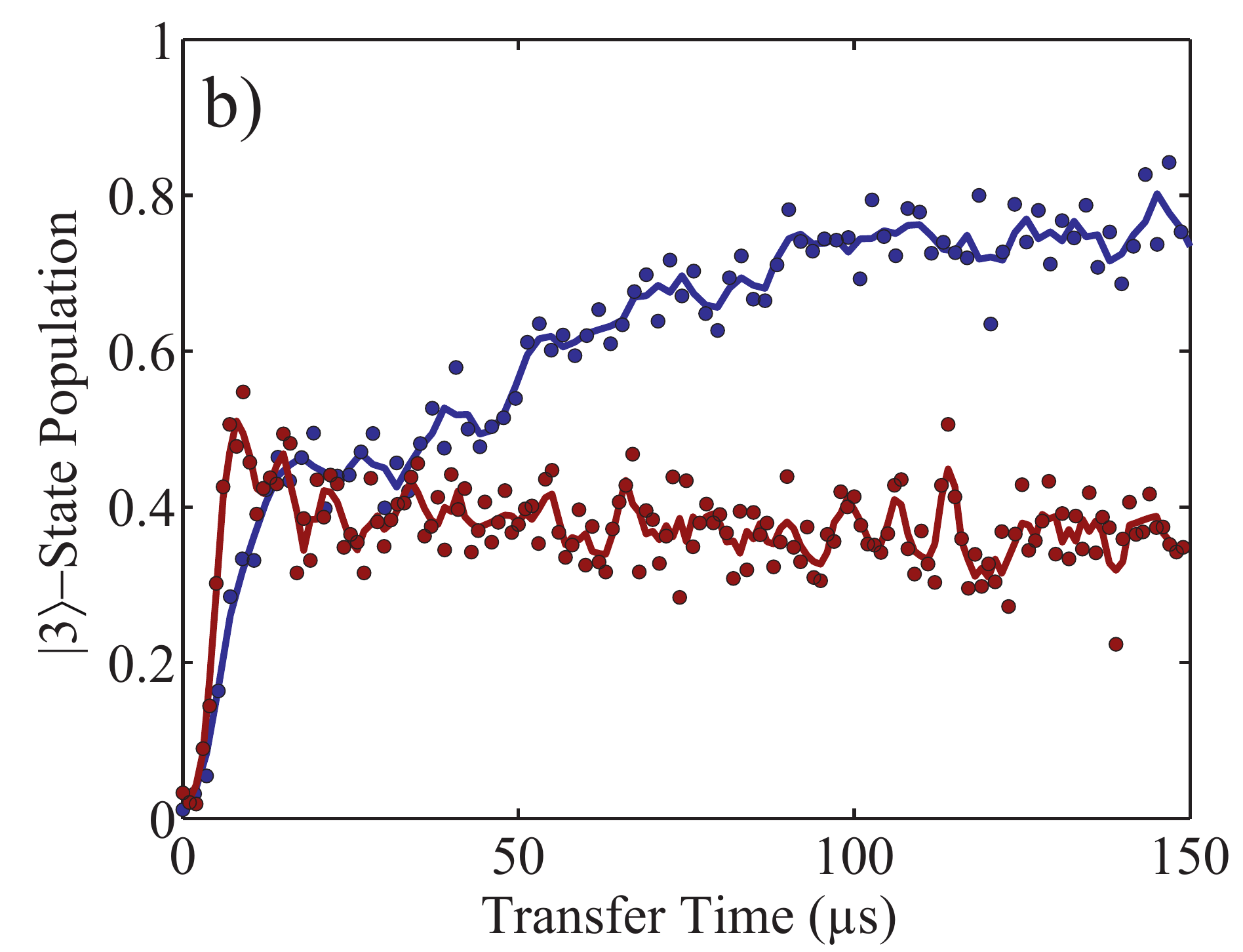}
	\label{fig:car_comp_stirap_rabi}
\caption[Population transfer dynamics of Rabi and STIRAP for a thermal motional state.]{\textbf{Population transfer dynamics of Rabi and STIRAP for a thermal motional state.} The transfer efficiency is displayed for a) the blue sideband and b) the carrier transition with the ion initialized by Doppler cooling. The delay scaling factor was chosen to be $s=0.5$ for sideband transitions and s=0.7 for carrier transitions. The lines are moving averages of the data and are guides to the eye. For clarity the error bars are omitted.}
\label{fig:comparison_Rabi_STIRAP}
\end{figure}
As expected, \cref{fig:comparison_Rabi_STIRAP} shows that the state evolution for Raman Rabi oscillations is faster than the corresponding STIRAP signal. After approximately $\unit{12}{\mu s}$ the maximum transfer efficiency of below $\unit{80}{\%}$ is reached for the blue sideband transition (\cref{fig:comparison_Rabi_STIRAP}a)). This transfer is sensitive to system parameters, especially to variations in Rabi frequency. The STIRAP transfer, however, is slower but reaches a transfer efficiency of higher than $\unit{85}{\%}$ for transfer times on the order of $\unit{150}{\mu s}$. It is worthwhile mentioning that the STIRAP transfer is limited by off-resonant excitation, which can be circumvented using a larger Raman detuning, whereas the transfer using Rabi oscillations is fundamentally limited by the different couplings between the motional states. On the carrier, the stronger motional state dependence of the Rabi frequency leads to a faster dephasing of the Rabi oscillations, as can be seen in \cref{fig:comparison_Rabi_STIRAP}b). Therefore, the transfer efficiency using Raman Rabi oscillations is reduced to $\unit{50}{\%}$ whereas the transfer using the STIRAP technique is still on the order of $\unit{75}{\%}$ for our system parameters. The transfer efficiency on a carrier transitions is reduced compared to the blue sideband transition since the Rabi frequency has a zero crossing at a motional state of $n=15$. Population in a range around this state can not be transferred efficiently by either technique.

\section{\label{sec:level6} Conclusion}

We have investigated the STIRAP technique to transfer population between two hyperfine states of a \Mg{25} ion. A systematic study of the transfer efficiency on the carrier and motional sidebands was performed for different pulse lengths and pulse delays with the ion initialized in the ground state of motion. Good agreement was found with numerical simulations. The insensitivity of STIRAP to the exact Rabi frequency was exploited to perform population transfer in the presence of an inherent range of Rabi frequencies found e.g. for thermally populated motional states. We demonstrated efficient population transfer on carrier and blue sideband transitions for Fock and thermal states. Experimentally the transfer was limited to $\sim 85~\%$ by off-resonant excitation to states not involved in the STIRAP process. However, this is not a fundamental limitation and could be overcome using a Raman laser system with a larger detuning, allowing transfer efficiencies approaching \unit{100}{\%}. 
In contrast, population transfer using Raman Rabi oscillations was shown to be faster, but less efficient for thermal motional states. 
We used the difference in blue and red sideband STIRAP transfer efficiency to detect the motional ground state population, from which a temperature in agreement with Doppler cooling temperature was extracted. This ground state population detection technique is robust against fluctuations of the Rabi frequencies and does not depend on the specific motional distribution. More elaborate pulse sequences using a series of STIRAP red and blue sidebands would even allow for the measurement of higher motional state populations, enhancing the accuracy of the determined temperature measurement. The same approach can be employed to determine the full motional state distribution \cite{muller_optimal_2015} or prepare strongly entangled states \cite{linington_robust_2008, noguchi_generation_2012}.

Another important application of the presented technique is the detection of small forces via excitation of a normal mode in a trapped ion crystal. Here, it has the potential to enhance the sensitivity of relative mass measurements \cite{drewsen_nondestructive_2004}, indirect internal state detection \cite{hume_trapped-ion_2011, wolf_quantum_2015} and the detection of small electrical \cite{biercuk_ultrasensitive_2010, narayanan_electric_2011} and optical \cite{clark_detection_2010, biercuk_phase-coherent_2011, lin_resonant_2013} forces. We  demonstrated the power of STIRAP in photon recoil spectroscopy \cite{wan_precision_2014, gebert_precision_2015} where the small force imprinted onto a two-ion crystal during absorption of a few photons leaves the motional state of the ions distributed over several trap levels. A STIRAP pulse on the red sideband probes the residual motional ground state population, which represents the spectroscopy signal.


\section{Acknowledgements}
We acknowledge the support of DFG through QUEST and Grant SCHM2678/3-1. This work was financially supported by the State of Lower-Saxony, Hannover, Germany. Y.W. acknowledges support from the Braunschweig International Graduate School of Metrology (B-IGSM). We thank Ian D. Leroux for stimulating discussions.

\section{References}
\bibliographystyle{h-physrev3}
\bibliography{bib}

\begin{thebibliography}{10}

\bibitem{leibfried_experimental_2003}
D.~Leibfried, B.~DeMarco, V.~Meyer, D.~Lucas, M.~Barrett, J.~Britton, W.~M.
  Itano, B.~Jelenkovi{\'c}, C.~Langer, T.~Rosenband, and D.~J. Wineland,
\newblock {\em Experimental demonstration of a robust, high-fidelity geometric
  two ion-qubit phase gate},
\newblock Nature {\bf 422}, 412 (2003).

\bibitem{haffner_quantum_2008}
H.~H{\"a}ffner, C.~Roos, and R.~Blatt,
\newblock {\em Quantum computing with trapped ions},
\newblock Phys. Rep. {\bf 469}, 155 (2008).

\bibitem{blatt_entangled_2008}
R.~Blatt and D.~Wineland,
\newblock {\em Entangled states of trapped atomic ions},
\newblock Nature {\bf 453}, 1008 (2008).

\bibitem{wineland_nobel_2013}
D.~J. Wineland,
\newblock {\em Nobel {Lecture}: {Superposition}, entanglement, and raising
  {Schr{\"o}dinger}{\textquoteright}s cat},
\newblock Rev. Mod. Phys. {\bf 85}, 1103 (2013).

\bibitem{schaetz_focus_2013}
T.~Schaetz, C.~R. Monroe, and T.~Esslinger,
\newblock {\em Focus on quantum simulation},
\newblock New J. Phys. {\bf 15}, 085009 (2013).

\bibitem{blatt_quantum_2012}
R.~Blatt and C.~F. Roos,
\newblock {\em Quantum simulations with trapped ions},
\newblock Nat Phys {\bf 8}, 277 (2012).

\bibitem{schmidt_spectroscopy_2005}
P.~O. Schmidt, T.~Rosenband, C.~Langer, W.~M. Itano, J.~C. Bergquist, and D.~J.
  Wineland,
\newblock {\em Spectroscopy {Using} {Quantum} {Logic}},
\newblock Science {\bf 309}, 749 (2005).

\bibitem{rosenband_frequency_2008}
T.~Rosenband, D.~B. Hume, P.~O. Schmidt, C.~W. Chou, A.~Brusch, L.~Lorini,
  W.~H. Oskay, R.~E. Drullinger, T.~M. Fortier, J.~E. Stalnaker, S.~A. Diddams,
  W.~C. Swann, N.~R. Newbury, W.~M. Itano, D.~J. Wineland, and J.~C. Bergquist,
\newblock {\em Frequency {Ratio} of {Al}$^{\textrm{+}}$ and {Hg}$^{\textrm{+}}$
  {Single}-{Ion} {Optical} {Clocks}; {Metrology} at the 17$^{\textrm{th}}$
  {Decimal} {Place}},
\newblock Science {\bf 319}, 1808 (2008).

\bibitem{hempel_entanglement-enhanced_2013}
C.~Hempel, B.~P. Lanyon, P.~Jurcevic, R.~Gerritsma, R.~Blatt, and C.~F. Roos,
\newblock {\em Entanglement-enhanced detection of single-photon scattering
  events},
\newblock Nat Photon {\bf 7}, 630 (2013).

\bibitem{huntemann_improved_2014}
N.~Huntemann, B.~Lipphardt, C.~Tamm, V.~Gerginov, S.~Weyers, and E.~Peik,
\newblock {\em Improved {Limit} on a {Temporal} {Variation} of
  m$_{\textrm{p}}$/m$_{\textrm{e}}$ from {Comparisons} of {Yb}$^{\textrm{+}}$
  and {Cs} {Atomic} {Clocks}},
\newblock Phys. Rev. Lett. {\bf 113}, 210802 (2014).

\bibitem{wan_precision_2014}
Y.~Wan, F.~Gebert, J.~B. W{\"u}bbena, N.~Scharnhorst, S.~Amairi, I.~D. Leroux,
  B.~Hemmerling, N.~L{\"o}rch, K.~Hammerer, and P.~O. Schmidt,
\newblock {\em Precision spectroscopy by photon-recoil signal amplification},
\newblock Nat Commun {\bf 5}, 4096 (2014).

\bibitem{gebert_precision_2015}
F.~Gebert, Y.~Wan, F.~Wolf, C.~N. Angstmann, J.~C. Berengut, and P.~O. Schmidt,
\newblock {\em Precision {Isotope} {Shift} {Measurements} in {Calcium} {Ions}
  {Using} {Quantum} {Logic} {Detection} {Schemes}},
\newblock Phys. Rev. Lett. {\bf 115}, 053003 (2015).

\bibitem{riebe_process_2006}
M.~Riebe, K.~Kim, P.~Schindler, T.~Monz, P.~Schmidt, T.~K{\"o}rber,
  W.~H{\"a}nsel, H.~H{\"a}ffner, C.~Roos, and R.~Blatt,
\newblock {\em Process {Tomography} of {Ion} {Trap} {Quantum} {Gates}},
\newblock Phys. Rev. Lett. {\bf 97}, 220407 (2006).

\bibitem{benhelm_towards_2008}
J.~Benhelm, G.~Kirchmair, C.~F. Roos, and R.~Blatt,
\newblock {\em Towards fault-tolerant quantum computing with trapped ions},
\newblock Nat Phys {\bf 4}, 463 (2008).

\bibitem{kirchmair_deterministic_2009}
G.~Kirchmair, J.~Benhelm, F.~Z{\"a}hringer, R.~Gerritsma, C.~F. Roos, and
  R.~Blatt,
\newblock {\em Deterministic entanglement of ions in thermal states of motion},
\newblock New J. Phys. {\bf 11}, 023002 (2009).

\bibitem{levitt_nmr_1979}
M.~H. Levitt and R.~Freeman,
\newblock {\em {NMR} population inversion using a composite pulse},
\newblock J. Magn. Reson. (1969) {\bf 33}, 473 (1979).

\bibitem{levitt_composite_1986}
M.~H. Levitt,
\newblock {\em Composite pulses},
\newblock Prog. Nucl. Magn. Reson. Spectrosc. {\bf 18}, 61 (1986).

\bibitem{vandersypen_nmr_2005}
L.~M.~K. Vandersypen and I.~L. Chuang,
\newblock {\em {NMR} techniques for quantum control and computation},
\newblock Rev. Mod. Phys. {\bf 76}, 1037 (2005).

\bibitem{gulde_implementation_2003}
S.~Gulde, M.~Riebe, G.~P.~T. Lancaster, C.~Becher, J.~Eschner, H.~H{\"a}ffner,
  F.~Schmidt-Kaler, I.~L. Chuang, and R.~Blatt,
\newblock {\em Implementation of the {Deutsch}-{Jozsa} algorithm on an ion-trap
  quantum computer},
\newblock Nature {\bf 421}, 48 (2003).

\bibitem{schmidt-kaler_realization_2003}
F.~Schmidt-Kaler, H.~H{\"a}ffner, M.~Riebe, S.~Gulde, G.~P.~T. Lancaster,
  T.~Deuschle, C.~Becher, C.~F. Roos, J.~Eschner, and R.~Blatt,
\newblock {\em Realization of the {Cirac}-{Zoller} controlled-{NOT} quantum
  gate},
\newblock Nature {\bf 422}, 408 (2003).

\bibitem{timoney_error-resistant_2008}
N.~Timoney, V.~Elman, S.~Glaser, C.~Weiss, M.~Johanning, W.~Neuhauser, and
  C.~Wunderlich,
\newblock {\em Error-resistant single-qubit gates with trapped ions},
\newblock Phys. Rev. A {\bf 77}, 052334 (2008).

\bibitem{ivanov_high-fidelity_2011}
S.~S. Ivanov and N.~V. Vitanov,
\newblock {\em High-fidelity local addressing of trapped ions and atoms by
  composite sequences of laser pulses},
\newblock Opt. Lett. {\bf 36}, 1275 (2011).

\bibitem{shappert_spatially_2013}
C.~M. Shappert, J.~T. Merrill, K.~R. Brown, J.~M. Amini, C.~Volin, S.~C. Doret,
  H.~Hayden, C.-S. Pai, K.~R. Brown, and A.~W. Harter,
\newblock {\em Spatially uniform single-qubit gate operations with near-field
  microwaves and composite pulse compensation},
\newblock New J. Phys. {\bf 15}, 083053 (2013).

\bibitem{mount_error_2015}
E.~Mount, C.~Kabytayev, S.~Crain, R.~Harper, S.-Y. Baek, G.~Vrijsen,
  S.~Flammia, K.~R. Brown, P.~Maunz, and J.~Kim,
\newblock {\em Error {Compensation} of {Single}-{Qubit} {Gates} in a {Surface}
  {Electrode} {Ion} {Trap} {Using} {Composite} {Pulses}},
\newblock arXiv:1504.01440 [quant-ph]  (2015).

\bibitem{bergmann_coherent_1998}
K.~Bergmann, H.~Theuer, and B.~W. Shore,
\newblock {\em Coherent population transfer among quantum states of atoms and
  molecules},
\newblock Rev. Mod. Phys. {\bf 70}, 1003 (1998).

\bibitem{bergmann_perspective:_2015}
K.~Bergmann, N.~V. Vitanov, and B.~W. Shore,
\newblock {\em Perspective: {Stimulated} {Raman} adiabatic passage: {The}
  status after 25 years},
\newblock J. Chem. Phys. {\bf 142}, 170901 (2015).

\bibitem{wunderlich_robust_2007}
C.~Wunderlich, T.~Hannemann, T.~K{\"o}rber, H.~H{\"a}ffner, C.~Roos,
  W.~H{\"a}nsel, R.~Blatt, and F.~Schmidt-Kaler,
\newblock {\em Robust state preparation of a single trapped ion by adiabatic
  passage},
\newblock J. Mod. Opt. {\bf 54}, 1541 (2007).

\bibitem{yamazaki_robust_2008}
R.~Yamazaki, K.-i. Kanda, F.~Inoue, K.~Toyoda, and S.~Urabe,
\newblock {\em Robust generation of superposition states},
\newblock Phys. Rev. A {\bf 78}, 023808 (2008).

\bibitem{noel_adiabatic_2012}
T.~Noel, M.~R. Dietrich, N.~Kurz, G.~Shu, J.~Wright, and B.~B. Blinov,
\newblock {\em Adiabatic passage in the presence of noise},
\newblock Phys. Rev. A {\bf 85}, 023401 (2012).

\bibitem{poschinger_interaction_2012}
U.~Poschinger, A.~Walther, M.~Hettrich, F.~Ziesel, and F.~Schmidt-Kaler,
\newblock {\em Interaction of a laser with a qubit in thermal motion and its
  application to robust and efficient readout},
\newblock Appl. Phys. B {\bf 107}, 1159 (2012).

\bibitem{watanabe_sideband_2011}
T.~Watanabe, S.~Nomura, K.~Toyoda, and S.~Urabe,
\newblock {\em Sideband excitation of trapped ions by rapid adiabatic passage
  for manipulation of motional states},
\newblock Phys. Rev. A {\bf 84}, 033412 (2011).

\bibitem{linington_robust_2008}
I.~E. Linington, P.~A. Ivanov, N.~V. Vitanov, and M.~B. Plenio,
\newblock {\em Robust control of quantized motional states of a chain of
  trapped ions by collective adiabatic passage},
\newblock Phys. Rev. A {\bf 77}, 063837 (2008).

\bibitem{toyoda_generation_2011}
K.~Toyoda, T.~Watanabe, T.~Kimura, S.~Nomura, S.~Haze, and S.~Urabe,
\newblock {\em Generation of {Dicke} states using adiabatic passage},
\newblock Phys. Rev. A {\bf 83}, 022315 (2011).

\bibitem{sorensen_efficient_2006}
J.~L. S{\o}rensen, D.~M{\o}ller, T.~Iversen, J.~B. Thomsen, F.~Jensen,
  P.~Staanum, D.~Voigt, and M.~Drewsen,
\newblock {\em Efficient coherent internal state transfer in trapped ions using
  stimulated {Raman} adiabatic passage},
\newblock New J. Phys. {\bf 8}, 261 (2006).

\bibitem{noguchi_generation_2012}
A.~Noguchi, K.~Toyoda, and S.~Urabe,
\newblock {\em Generation of {Dicke} {States} with {Phonon}-{Mediated}
  {Multilevel} {Stimulated} {Raman} {Adiabatic} {Passage}},
\newblock Phys. Rev. Lett. {\bf 109}, 260502 (2012).

\bibitem{moller_efficient_2007}
D.~M{\o}ller, J.~L. S{\o}rensen, J.~B. Thomsen, and M.~Drewsen,
\newblock {\em Efficient qubit detection using alkaline-earth-metal ions and a
  double stimulated {Raman} adiabatic process},
\newblock Phys. Rev. A {\bf 76}, 062321 (2007).

\bibitem{kamsap_coherent_2013}
M.~R. Kamsap, T.~B. Ekogo, J.~Pedregosa-Gutierrez, G.~Hagel, M.~Houssin,
  O.~Morizot, M.~Knoop, and C.~Champenois,
\newblock {\em Coherent internal state transfer by a three-photon {STIRAP}-like
  scheme for many-atom samples},
\newblock J. Phys. B: At. Mol. Opt. Phys. {\bf 46}, 145502 (2013).

\bibitem{wineland_experimental_1998}
D.~J. Wineland, C.~Monroe, W.~M. Itano, D.~Leibfried, B.~E. King, and D.~M.
  Meekhof,
\newblock {\em Experimental issues in coherent quantum-state manipulation of
  trapped atomic ions},
\newblock J. Res. Natl. Inst. Stand. Technol. {\bf 103}, 259 (1998).

\bibitem{wan_efficient_2015}
Y.~Wan, F.~Gebert, F.~Wolf, and P.~O. Schmidt,
\newblock {\em Efficient sympathetic motional-ground-state cooling of a
  molecular ion},
\newblock Phys. Rev. A {\bf 91}, 043425 (2015).

\bibitem{fewell_coherent_1997}
M.~P. Fewell, B.~W. Shore, and K.~Bergmann,
\newblock {\em Coherent {Population} {Transfer} among {Three} {States}: {Full}
  {Algebraic} {Solutions} and the {Relevance} of {Non} {Adiabatic} {Processes}
  to {Transfer} by {Delayed} {Pulses}},
\newblock Aust. J. Phys. {\bf 50}, 281 (1997).

\bibitem{klein_robust_2007}
J.~Klein, F.~Beil, and T.~Halfmann,
\newblock {\em Robust {Population} {Transfer} by {Stimulated} {Raman}
  {Adiabatic} {Passage} in a
  {Pr}$^{\textrm{3+}}$:{Y}$_{\textrm{2}}${SiO}$_{\textrm{5}}$ {Crystal}},
\newblock Phys. Rev. Lett. {\bf 99}, 113003 (2007).

\bibitem{born_beweis_1928}
M.~Born and V.~Fock,
\newblock {\em Beweis des {Adiabatensatzes}},
\newblock Z. Physik {\bf 51}, 165 (1928).

\bibitem{gaubatz_population_1990}
U.~Gaubatz, P.~Rudecki, S.~Schiemann, and K.~Bergmann,
\newblock {\em Population transfer between molecular vibrational levels by
  stimulated {Raman} scattering with partially overlapping laser fields. {A}
  new concept and experimental results},
\newblock J. Chem. Phys. {\bf 92}, 5363 (1990).

\bibitem{hemmerling_single_2011}
B.~Hemmerling, F.~Gebert, Y.~Wan, D.~Nigg, I.~V. Sherstov, and P.~O. Schmidt,
\newblock {\em A {Single} {Laser} {System} for {Ground} {State} {Cooling} of
  $^{\textrm{25}}${Mg}$^{\textrm{+}}$},
\newblock Appl. Phys. B {\bf 104}, 583 (2011).

\bibitem{hemmerling_novel_2012}
B.~Hemmerling, F.~Gebert, Y.~Wan, and P.~O. Schmidt,
\newblock {\em A novel, robust quantum detection scheme},
\newblock New J. Phys. {\bf 14}, 023043 (2012).

\bibitem{pham_general-purpose_2005}
P.~T.~T. Pham,
\newblock {\em A general-purpose pulse sequencer for quantum computing},
\newblock {PhD} {Thesis}, Massachusetts Institute of Technology, Cambridge,
  Massachusetts, USA, (2005).

\bibitem{schindler_frequency_2008}
P.~Schindler,
\newblock {\em Frequency synthesis and pulse shaping for quantum information
  processing with trapped ions},
\newblock {PhD} {Thesis}, University of Innsbruck, Innsbruck, Austria, (2008).

\bibitem{gebert_damage-free_2014}
F.~Gebert, M.~H. Frosz, T.~Weiss, Y.~Wan, A.~Ermolov, N.~Y. Joly, P.~O.
  Schmidt, and P.~S.~J. Russell,
\newblock {\em Damage-free single-mode transmission of deep-{UV} light in
  hollow-core {PCF}},
\newblock Opt. Express {\bf 22}, 15388 (2014).

\bibitem{tan_computational_1999}
S.~M. Tan,
\newblock {\em A computational toolbox for quantum and atomic optics},
\newblock J. Opt. B: Quantum Semiclass. Opt. {\bf 1}, 424 (1999).

\bibitem{matlab_version_2013}
{MATLAB},
\newblock {\em version 8.2.0.701 ({R}2013b)} (The MathWorks Inc., Natick,
  Massachusetts, 2013).

\bibitem{Poulsen_sideband_2011}
G.~Poulsen,
\newblock {\em Sideband {Cooling} of {Atomic} and {Molecular} {Ions}},
\newblock {PhD} {Thesis}, University of Aarhus, Aarhus, Denmark, (2011).

\bibitem{hemmerling_towards_2011}
B.~Hemmerling,
\newblock {\em Towards {Direct} {Frequency} {Comb} {Spectroscopy} using
  {Quantum} {Logic}},
\newblock {PhD} {Thesis}, Hannover, (2011).

\bibitem{muller_optimal_2015}
M.~M. M{\"u}ller, U.~G. Poschinger, T.~Calarco, S.~Montangero, and
  F.~Schmidt-Kaler,
\newblock {\em Optimal {Phonon}-to-{Spin} {Mapping} in a system of a trapped
  ion},
\newblock arXiv:1504.02858 [quant-ph]  (2015).

\bibitem{drewsen_nondestructive_2004}
M.~Drewsen, A.~Mortensen, R.~Martinussen, P.~Staanum, and J.~L. S{\o}rensen,
\newblock {\em Nondestructive {Identification} of {Cold} and {Extremely}
  {Localized} {Single} {Molecular} {Ions}},
\newblock Phys. Rev. Lett. {\bf 93}, 243201 (2004).

\bibitem{hume_trapped-ion_2011}
D.~B. Hume, C.~W. Chou, D.~R. Leibrandt, M.~J. Thorpe, D.~J. Wineland, and
  T.~Rosenband,
\newblock {\em Trapped-{Ion} {State} {Detection} through {Coherent} {Motion}},
\newblock Phys. Rev. Lett. {\bf 107}, 243902 (2011).

\bibitem{wolf_quantum_2015}
F.~Wolf, Y.~Wan, J.~C. Heip, F.~Gebert, C.~Shi, and P.~O. Schmidt,
\newblock {\em Quantum logic with molecular ions},
\newblock arXiv:1507.07511 [physics]  (2015).

\bibitem{biercuk_ultrasensitive_2010}
M.~J. Biercuk, H.~Uys, J.~W. Britton, A.~P. VanDevender, and J.~J. Bollinger,
\newblock {\em Ultrasensitive detection of force and displacement using trapped
  ions},
\newblock Nat Nano {\bf 5}, 646 (2010).

\bibitem{narayanan_electric_2011}
S.~Narayanan, N.~Daniilidis, S.~A. M{\"o}ller, R.~Clark, F.~Ziesel, K.~Singer,
  F.~Schmidt-Kaler, and H.~H{\"a}ffner,
\newblock {\em Electric field compensation and sensing with a single ion in a
  planar trap},
\newblock J. Appl. Phys. {\bf 110}, 114909 (2011).

\bibitem{clark_detection_2010}
C.~R. Clark, J.~E. Goeders, Y.~K. Dodia, C.~R. Viteri, and K.~R. Brown,
\newblock {\em Detection of single-ion spectra by {Coulomb}-crystal heating},
\newblock Phys. Rev. A {\bf 81}, 043428 (2010).

\bibitem{biercuk_phase-coherent_2011}
M.~J. Biercuk, H.~Uys, J.~W. Britton, A.~P. VanDevender, and J.~J. Bollinger,
\newblock {\em Phase-coherent detection of an optical dipole force by {Doppler}
  velocimetry},
\newblock Opt. Express {\bf 19}, 10304 (2011).

\bibitem{lin_resonant_2013}
Y.-W. Lin, S.~Williams, and B.~C. Odom,
\newblock {\em Resonant few-photon excitation of a single-ion oscillator},
\newblock Phys. Rev. A {\bf 87}, 011402 (2013).

\end{thebibliography}

\end{document}